\documentclass[aps,pra,twocolumn,showpacs, superscriptaddress,eqsecnum,10pt,floatfix]{revtex4-1}

\usepackage{placeins}
\usepackage{amsfonts}
\usepackage{amssymb}
\usepackage{amsmath, mathtools}
\usepackage{graphicx} 
\usepackage{datetime}
\usepackage{color}
\usepackage{enumerate}
\usepackage{pifont,array}
\usepackage{braket}

\newcommand{\di}[1]{\frac{d}{d#1}}

\newcommand{\expectn}[1]{\langle#1\rangle}

\begin{document}

\title{Extracting unambiguous information from a single qubit by sequential observers}

\author{Dov Fields}
\affiliation{Department of Physics and Astronomy, Hunter College of the City University of New York, 695 Park Avenue, New York, NY 10065, USA}
\affiliation{Physics Program, The Graduate Center of the City University of New York, 365 Fifth Avenue, New York, NY 10016, USA}
\author{Rui Han}
\affiliation{Max Planck Institute for the Science of Light, 91058 Erlangen, Germany}
\affiliation{Institute of Optics, Information and Photonics, University of Erlangen-N\"urnberg, 91058 Erlangen, Germany}
\author{Mark Hillery}
\affiliation{Department of Physics and Astronomy, Hunter College of the City University of New York, 695 Park Avenue, New York, NY 10065, USA}
\affiliation{Physics Program, The Graduate Center of the City University of New York, 365 Fifth Avenue, New York, NY 10016, USA}
\author{J\'anos A. Bergou}
\affiliation{Department of Physics and Astronomy, Hunter College of the City University of New York, 695 Park Avenue, New York, NY 10065, USA}
\affiliation{Physics Program, The Graduate Center of the City University of New York, 365 Fifth Avenue, New York, NY 10016, USA}

\begin{abstract}
In a recent paper [\prl \ {\bf 111}, 100501 (2013)], a scheme was proposed where subsequent observers can extract unambiguous information about the initial state of a qubit, with finite joint probability of success. 
Here, we generalize the problem for arbitrary preparation probabilities (arbitrary priors). We discuss two different schemes: one where only the joint probability of success is maximized and another where, in addition, the joint probability of failure is also minimized. We also derive the mutual information for these schemes and show that there are some parameter regions for the scheme without minimizing the joint failure probability where, even though the joint success probability is maximum, no information is actually transmitted by Alice.
\end{abstract}

\maketitle

\section{Introduction}

Discrimination procedures for distinguishing between quantum states have been a topic of much interest with many applications in,  e.g., the area of secure distribution of information. The laws of quantum mechanics rule out perfect discrimination of non-orthogonal quantum states. For recent reviews of quantum state discrimination, see \cite{Barnettreview,Bergoureview,Baereview}. One discrimination procedure is unambiguous discrimination (UD). Posed initially by Ivanovic, Dieks, and Peres \cite{Ivanovic1987,Dieks1988,Peres1988}, the goal of this procedure is to identify the given state with no error. Doing this requires setting up a measurement such that there is some possibility for the measurement to fail, in which case no conclusion is drawn, and some possibility that the states are identified with no error. The question of how to maximize one's probability of success for discriminating between two non-orthogonal states was solved completely for arbitrary priors by Jaeger and Shimony \cite{Jaeger1995}. 

As an interesting extension of UD by a single observer, a scheme for sequential unambiguous state discrimination by multiple observers was proposed in Ref. \cite{Bergou2013}. In this work a communication protocol was introduced among three parties - Alice, Bob, and Charlie - where Alice prepares a qubit in one of two non-orthogonal quantum states, and passes it to Bob. The states are also known to Bob and Charlie, they just don't know which state has actually been prepared. After performing a measurement to discriminate between these two states, Bob sends the qubit to Charlie so that he can also, independently, have some probability of learning Alice's initial state, provided Bob did not extract all of the available unambiguous information from the qubit with his measurement. 

The paper demonstrated that one can get around the constraints of the no-broadcasting theorem \cite{Barnum96} and the collapse postulate \cite{Neumann32}, in a probabilistic manner, i.e., with a finite probability of success. These protocols are intrinsically related to the no-cloning theorem \cite{Wootters82,Dieks82} that forbids deterministic cloning but allows probabilistically perfect cloning with a finite ($<1$) probability of success \cite{Duan98,Yerokhin16}. 

The model introduced in \cite{Bergou2013} showed that sequential unambiguous discrimination is possible. It also presented a solution for the case of equal priors where each of the two observers, Bob and Charlie, can identify the two states with finite probability of success. In a subsequent paper, a different solution to the same problem was found \cite{Pang2013}. It showed that for some range of the parameters, maximum joint probability of success is achieved when Bob and Charlie choose their measurements such that, in the case of success, they always identify the same state and never the other. We also note that soon after the initial publication \cite{Bergou2013}, the scheme has been verified experimentally \cite{Solis2016} and extended from discrete to continuous variable states \cite{Namkung2018}.

The goal of this paper is to generalize the sequential unambiguous discrimination scheme to arbitrary priors and to present a complete discussion of the general case. In order to extend the usefulness of the solution found in Ref. \cite{Pang2013}, we adopt the so-called flip-flop measurement \cite{Bennett92} for the present scheme, and determine the optimal flipping rate. We also consider these schemes from an information theoretical view point and derive the mutual information between Alice and Bob (same as between Alice and Charlie in the optimal case). Mutual information is a standard quantity that quantifies channel capacity, or how much information can be transmitted between the involved parties \cite{Cover2006}.  Mutual information for the non-sequential UD scheme has been calculated previously \cite{Peres2002} and we extend the calculation here to the sequential scheme. We show that, while the standard mutual information is maximized for the set-up that always identifies only one of the states when it succeeds, the so-called conditional mutual information that is optimized under the condition that the resulting bit-string is maximally unbiased, is better suited for discussing  application to quantum communication. 

The paper is organized as follows. We present a brief overview of sequential UD for equal priors in Sec. II, and discuss the range of validity of the solutions found in \cite{Bergou2013} and \cite{Pang2013}. We also propose a multiparty communication (QKD) protocol, based on the scheme. Section III presents the full analytical theory of the simultaneous optimization of the joint probability of success (maximum) and joint probability of failure (minimum) for general priors. Section IV discusses the optimization of the joint probability of success with no simultaneous minimization of the joint failure probability. In Sec. V, we present the flip-flop measurement. In Sec. VI, we derive the mutual information between the various communicating parties. We conclude with a brief discussion of the results.

\section{A brief overview of sequential unambiguous discrimination}

In the sequential unambiguous discrimination scheme, Alice prepares a qubit in one of two non-orthogonal states, either $\ket{\psi_{1}}$ or $\ket{\psi_{2}}$. The probability (prior probability or, simply, prior) that $\ket{\psi_{i}}$ is prepared is $\eta_{i}$ $\{i=1,2\}$, such that  $\eta_{1} + \eta_{2} = 1$, so one of the states is always prepared. In the original work \cite{Bergou2013}, only the case $\eta_{1} = \eta_{2} = 1/2$ was considered. Here we address the problem with arbitrary priors and other generalizations. 

The states and their priors are also known to Bob and Charlie, they just do not know which state the qubit was actually prepared in. After the preparation Alice sends the qubit to Bob, the first observer in the sequence, who performs a measurement on the qubit, possibly a POVM, and sends the qubit, he just measured, to Charlie, who then also performs a measurement (POVM) on the qubit he received. The goal for them is to maximize their joint probability of succeeding. This goal is compatible with additional optimizations and in what follows we will analyze these options in detail.

In the original treatment the standard POVM formalism was employed. For the purposes of the present work we find the alternative but equivalent formalism, based on Neumark's extension, more suitable \cite{Neumark1943}. A more accessible and tutorial treatment of the Neumark method can be found in, e.g., Ref. \cite{BH2013}. In this formalism one first entangles the qubit with an ancilla and then performs a standard projective measurement on the ancillary system. The interaction of the qubit with the ancilla is described by a unitary time evolution operator,
\begin{eqnarray}
	U_{\text{b}}\ket{\psi_{1}}\ket{i} &=& \sqrt{p_{1\text{b}}}\ket{\varphi_{1}}\ket{1} + \sqrt{q_{1\text{b}}}\ket{\Phi_{1}}\ket{0}, \\
	U_{\text{b}}\ket{\psi_{2}}\ket{i} &=&  \sqrt{p_{2\text{b}}}\ket{\varphi_{2}}\ket{2} + \sqrt{q_{2\text{b}}}\ket{\Phi_{2}}\ket{0}.
	\label{BobSequentialNeumark}
\end{eqnarray}
Here the subscript $b$ stands for Bob, $|i\rangle$ is the initial state of the ancilla while $|0\rangle$, $|1\rangle$ and $|2\rangle$ are three orthogonal states of the ancillary system.  If Bob performs a measurement on the ancilla in the basis formed by these three states and finds either $|1\rangle$ or $|2\rangle$ as the measurement outcome, he will know what state Alice has prepared. If, on the other hand, he finds $|0\rangle$ as the outcome of his measurement, he will not acquire unambiguous information about the input state, hence this result is inconclusive. Therefore, $p_{i\text{b}}$ is Bob's success probability of unambiguously identifying the input state $|\psi_{i}\rangle$ and $q_{i\text{b}}$ is Bob's probability of failing to identify the input state.  $\ket{\varphi_{i}}$ and $\ket{\Phi_{i}}$ ($i=1,2$) are the post-measurement states of the qubit associated with the various outcomes of the measurement performed on the ancilla.

After Bob has performed his state-identifying measurement, he passes the qubit to Charlie, whose task is also to unambiguously identify the initial state of the qubit that Alice prepared. It is known that for unambiguous identification the states to be identified must be linearly independent \cite{Chefles98}. For a qubit, this means that two pure states can be unambiguously discriminated. This requirement puts serious restrictions on how Bob can design the post-measurement states. There is one additional requirement. We also want that the post-measurement states of the qubit carry no information about the outcome of Bob's measurement, a condition that is central to applications for quantum communication.   

The choice $\ket{\varphi_{i}} = \ket{\Phi_{i}}$, made in \cite{Bergou2013}, is mandated by these requirements. It ensures that Charlie receives the pure state $\ket{\varphi_{1}}$ ($\ket{\varphi_{2}}$) if Alice sent $\ket{\psi_{1}}$ ($\ket{\psi_{2}}$). Charlie receives these states independently of whether Bob's measurement succeeded or failed. In order to learn what Alice sent, Charlie's task is to unambiguously discriminate between $\ket{\varphi_{1}}$ and $\ket{\varphi_{2}}$. His measurement, again employing the Neumark method, can be represented as

\begin{eqnarray}
	U_{\text{c}}\ket{\varphi_{1}}\ket{i} &=& \sqrt{p_{1\text{c}}}\ket{\theta_{1}}\ket{1} + \sqrt{q_{1\text{c}}}\ket{\Theta_{1}}\ket{0},\\
	U_{\text{c}}\ket{\varphi_{2}}\ket{i} &=& \sqrt{p_{2\text{c}}}\ket{\theta_{2}}\ket{2} + \sqrt{q_{2\text{c}}}\ket{\Theta_{2}}\ket{0}.
	\label{CharlieSequentialNeumark}
\end{eqnarray}

It has been shown previously that, in order to optimally discriminate between $\ket{\varphi_{1}}$ and $\ket{\varphi_{2}}$, Charlie must choose $\ket{\Theta_{1}} = \ket{\Theta_{2}} \equiv \ket{\theta_{0}}$ (see, e.g., \cite{Bergoureview}). In order to simplify the following discussion, we introduce the notation $\braket{\psi_{1}|\psi_{2}} = s$ and $\braket{\varphi_{1}|\varphi_{2}} = t$,  We can express the constraints, resulting from the unitarity of $U_{b}$ and $U_{c}$, in terms of these parameters as
\begin{equation}
	p_{\text{jb}} + q_{jb} = p_{jc} + q_{jc} = 1 
\label{BCnorm}
\end{equation}
for $j=1,2$ and
\begin{equation}
	\frac{s}{t} = \sqrt{q_{1\text{b}}q_{2\text{b}}}, \; t = \sqrt{q_{1\text{c}}q_{2\text{c}}}.
\label{BCconstraints}
\end{equation}

The average probability that both Bob and Charlie succeed in unambiguously identifying the state that Alice sent, the joint success probability, can be written as:
\begin{eqnarray}
	P_\mathrm{ss} &=& \eta_{1}p_{1\text{b}}p_{1\text{c}} + \eta_{2}p_{2\text{b}}p_{2\text{c}} 
	\label{JointSuccess1}
\end{eqnarray}
This is the central quantity for the rest of this work. The main goal is to optimize this expression under the constraints given by Eqs. \eqref{BCnorm} and \eqref{BCconstraints}.

By making use of the constraints given in Eqs. \eqref{BCnorm} and \eqref{BCconstraints}, we can write $P_{{\mathrm{ss}}}$ as
\begin{eqnarray}
	P_\mathrm{ss} &=& \eta_{1}\left( 1 - q_{1\text{b}} \right)\left( 1 - q_{1\text{c}} \right) \nonumber \\
	&& + \eta_{2}\left( 1 - \frac{s^{2}}{t^{2}q_{1\text{b}}} - \frac{t^{2}}{q_{1\text{c}}} + \frac{s^{2}}{q_{1\text{b}}q_{1\text{c}}} \right).
	\label{JointSuccess2}
\end{eqnarray}

Equations \eqref{BCnorm}-\eqref{JointSuccess2} represent the starting point for the various optimization schemes and discussions in the next four sections. In particular, Eq. \eqref{JointSuccess2} is a function of three independent parameters, $t$, $q_{1b}$ and $q_{1c}$. Their range is given by $s \leq t \leq 1$, $\frac{s^2}{t^2}\leq q_{1b} \leq 1$ and $t^2\leq q_{1c} \leq 1$. For the optimal $P_\mathrm{ss}$, the parameters are either internal points in these intervals or lie at the boundary. In the first case the derivatives of $P_{{\mathrm{ss}}}$ with respect to the variables $t$, $q_{1b}$ and $q_{1c}$ exist and the optimum can be found by setting the derivatives equal to 0. To find the global optimum, the boundary points need to be compared to the internal optimum points.

Before we move on to discuss the general case, we deal with the special case of $\eta_{1} = \eta_{2} = 1/2$, which was the case considered in Refs. \cite{Bergou2013} and \cite{Pang2013}. It was shown in \cite{Bergou2013} that $t^{2}=s$ for optimum joint probability of success. We will see in the next sections that this remains the optimal choice for general priors, as well. Under this condition, the equations \eqref{BCnorm}-\eqref{JointSuccess2} are completely symmetric in the indices 1 and 2, and also in $b$ and $c$. This immediately yields $q_{1b}=q_{1c}=q_{2b}=q_{2c}=\sqrt{s}$ for the internal point solution. Inserting these values into Eq. \eqref{JointSuccess2}, gives 
\begin{equation}
P_{ss,1}^{opt} = (1-\sqrt{s})^{2} 
\label{Pssopt1}
\end{equation}
for the optimum joint probability of success, which is the result found in \cite{Bergou2013}. For the boundary solution Bob can choose either $q_{1b}=1$ or $q_{2b}=1$ but not both. Let us assume $q_{1b}=1$, then Eq. \eqref{BCconstraints} leads to $q_{2b}=s$. Similarly, Charlie can choose either $q_{1c}=1$ or $q_{2c}=1$ but not both, for his boundary solution. If he chooses $q_{2c}=1$, he always fails to identify the second state and sometimes identifies the first. Since Bob always fails to identify the first state and sometimes identifies the second state, their joint probability of success is zero, giving the minimum of $P_{ss}$. So, Charlie must choose $q_{1c}=1$ leading to $q_{2b}=s$. Inserting these into Eq. \eqref{JointSuccess2}, yields
\begin{equation}
P_{ss,2}^{opt} = \frac{1}{2}(1-s)^{2} 
\label{Pssopt2}
\end{equation}
for the optimum joint probability of success, which is the result found in \cite{Pang2013}. As it turns out, $P_{ss,1}$ is optimum if $s \leq s_{c}$ and  $P_{ss,2}$ is optimum if $s > s_{c}$, where $s_{c} = (\sqrt{2}-1)^{2}$ is the critical value of the overlap parameter where the two solutions become equal.

Clearly, a two-state QKD protocol can be based on the sequential scheme. It is very closely related to the B'92 protocol \cite{Bennett92}, extending it to multiple recipients. Alice encodes the bit value 0 into the first state and 1 into the second state. She prepares a large number of qubits at random in one of these states and sends them to Bob who performs the above described state identifying measurement on them and sends the qubits in their post-measurement states to Charlie who performs an optimal UD measurement on them. They publicly announce the instances when they succeeded but not the result. They keep the results when they succeed and discard the rest.  Since Alice knows what she prepared in those instances, she will share a string of 0's and 1's with Bob in those instances when Bob succeeds and, similarly, a separate string with Charlie in the instances when Charlie succeeds. In addition, in the instances when both Bob and Charlie succeed, they will share a subset of their bit-strings that is common to all three of them. These bit strings serve as the raw key and the rest of the protocol (checking for the presence of eavesdropper(s) and distilling a communication key) follows the same lines as in the original B'92 protocol. The established communication keys can serve as secure keys for a secure three-way communication protocol. It is clear that for this QKD protocol the measurement presented in \cite{Bergou2013} has to employed. The measurement  presented in \cite{Pang2013} cannot be used in communication protocols since it generates a string of identical bit values, either all 0's or all 1's, which is clearly not what is needed for a key.

After these preliminaries, we now proceed to the discussion of the general case. At this point, we arrive at a juncture, one can follow one of two ways. One can maximize the joint probability of success and simultaneously minimize the joint probability of failure,
\begin{equation}
P_{\mathrm{ff}} = \eta_{1} q_{1b} q_{1c} + \eta_{2}  q_{2b} q_{2c} .
\label{Pff}
\end{equation}
Alternatively, one can maximize the joint probability of success, without minimizing the joint probability of failure. The two methods yield slightly different results. The first allows for a fully analytical treatment while the second can be treated numerically only. We present the first approach in the next section and the second in Sec. IV.

\section{Simultaneous optimization of the joint probability of success the joint probability of failure}

The joint probability of failure, Eq. \eqref{Pff}, can be optimized independently of the rest of the problem, based on the following observation. Taking the product of the constraints in Eq. \eqref{BCconstraints} yields $q_{1b} q_{1c} q_{2b} q_{2c} = s^{2}$, which is independent of $t$. So, $q_{2b} q_{2c}$ can be expressed in terms of of the failure probabilities of the first state,
\begin{equation}
q_{2b} q_{2c} = \frac{s^{2}}{q_{1b} q_{1c}} .
\label{qconstraint}
\end{equation}
Inserting this expression into Eq. \eqref{Pff}, $P_{\mathrm{ff}}$ will depend only on the single combination of the parameters, $q_{1b} q_{1c}$. The optimization with respect to this parameter is straightforward, with the result
\begin{eqnarray}
    q_{1b}^{opt} q_{1c}^{opt} = \left\{ \begin{array}{ll}
    \sqrt{\frac{\eta_{2}}{\eta_{1}}}s & \mbox{ if
    $\frac{s^{2}}{1+s^{2}} \leq \eta_{1}
    \leq  \frac{1}{1+s^{2}}$} \ , \\
    1 & \mbox{ if $\eta_{1} <
    \frac{s^{2}}{1+s^{2}} $} \ , \\ 
    s^{2} & \mbox{ if $\frac{1}{1+s^{2}}
    < \eta_{1}$} \ .
    \end{array}
    \right.
    \label{q1opt}
\end{eqnarray}
Substituting the optimal  values into Eq. \eqref{Pff} yields
\begin{eqnarray}
    \label{Qopt}
    P_{\mathrm{ff}}^{\mathrm{opt}} = \left\{ \begin{array}{ll}
    2\sqrt{\eta_{1}\eta_{2}}s & \mbox{ if
    $\frac{s^{2}}{1+s^{2}} \leq \eta_{1}
    \leq  \frac{1}{1+s^{2}}$} \ , \\
    \eta_{1} + \eta_{2} s^{2} & \mbox{ if $\eta_{1} <
    \frac{s^{2}}{1+s^{2}} $} \ , \\ 
    \eta_{2} + \eta_{1} s^{2} & \mbox{ if $\frac{1}{1+s^{2}}
    < \eta_{1}$} \ .
    \end{array}
    \right.
\end{eqnarray}
Interestingly, this expression is identical to the one obtained for optimal unambiguous discrimination of the two states by Bob alone \cite{Jaeger1995,Bergoureview}. This was to be expected, since Bob can first perform a partial discrimination of the two states and then in a second step a full discrimination of the remaining states, i.e., he can assume the role of Charlie in the sequence. What the above result tells us is that no matter in how many steps the discrimination is performed, its optimal failure probability is always given by the above equation. Thus, quantum mechanics sets a universal bound on the global failure probability.

The individual success probabilities of Bob and Charlie are, however, subject to further optimization.  In addition to Eqs. \eqref{BCnorm} and \eqref{BCconstraints}, we now have Eqs. \eqref{qconstraint} and \eqref{q1opt} as constraints for the optimization of $P_\mathrm{ss}$. From the first line in Eq. \eqref{q1opt}, we can express $q_{1c}$ in terms of $q_{1b}$, as
\begin{equation}
q_{1c}^{opt} =  \sqrt{\frac{\eta_{2}}{\eta_{1}}} \frac{s}{q_{1b}^{opt}} .
\end{equation}
Using this in Eq \eqref{JointSuccess2}, $P_\mathrm{ss}$ is a function of $t$ and $q_{1b}^{opt}$ only,
\begin{eqnarray}
	P_\mathrm{ss} &=& 1 + 2 \sqrt{\eta_{1}\eta_{2}}  - \eta_{1} q_{1\text{b}}  - \sqrt{\eta_{1}\eta_{2}} \frac{s}{q_{1\text{b}}} \nonumber \\
	&&  - \eta_{2} \frac{s^{2}}{t^{2}q_{1\text{b}}} - \sqrt{\eta_{1}\eta_{2}} \frac{t^{2}}{s} q_{1\text{b}} .
	\label{JointSuccess3}
\end{eqnarray}

The optimization with respect to $t$ and $q_{1b}^{opt}$ is again straightforward, yielding the unique solutions $t^{2} = s$ and 
\begin{eqnarray}
    q_{1b}^{opt}  = \left\{ \begin{array}{ll}
   (\frac{\eta_{2}}{\eta_{1}})^{1/4}\sqrt{s}& \mbox{ if
    $\frac{s^{2}}{1+s^{2}} \leq \eta_{1}
    \leq  \frac{1}{1+s^{2}}$} \ , \\
    1 & \mbox{ if $\eta_{1} <
    \frac{s^{2}}{1+s^{2}} $} \ , \\ 
    s & \mbox{ if $\frac{1}{1+s^{2}}
    < \eta_{1}$} \ .
    \end{array}
    \right.
    \label{q1optopt}
\end{eqnarray}

Using these expressions in Eq. \eqref{JointSuccess3} yields
\begin{widetext}
\begin{eqnarray}
    \label{Pssopt}
    P_{\mathrm{ss}}^{\mathrm{opt}} = \left\{ \begin{array}{ll}
    \left(\sqrt{\eta_{1}} - (\eta_{1}\eta_{2}\right)^{1/4}\sqrt{s})^{2} +  \left(\sqrt{\eta_{2}} - (\eta_{1}\eta_{2})^{1/4}\sqrt{s}\right)^{2} & \mbox{ if
    $\frac{s^{2}}{1+s^{2}} \leq \eta_{1}
    \leq  \frac{1}{1+s^{2}}$} \ , \\
    \eta_{2} (1-s)^{2} & \mbox{ if $\eta_{1} <
    \frac{s^{2}}{1+s^{2}} $} \ , \\ 
    \eta_{1} (1-s)^{2} & \mbox{ if $\frac{1}{1+s^{2}}
    < \eta_{1}$} \ .
    \end{array}
    \right.
\end{eqnarray}
\end{widetext}
for the optimal joint success probability, under the condition that the joint probability of failure is minimum. 

The solution is fully analytic and unique. The internal point solution, first line in Eq. \eqref{Pssopt}, coexists with the boundary solutions, second and third line, in various regimes of the priors. It is interesting to note that \eqref{Pssopt} reduces to \eqref{Pssopt1} for equal priors, $\eta_{1}=\eta_{2}=1/2$. The boundary solution \eqref{Pssopt2} is not compatible with the minimum joint probability of failure in a finite range of the prior probabilities.

Next, we relax the requirement that the joint probability of failure is at its minimum. The joint probability of success can still be optimized and we will present this case in the next section.

\section{Optimizing the joint probability of success without minimizing the joint probability of failure}

First, let us consider the extrema of the joint probability of success with respect to $t$. They require that $t$ is either on the boundary of the allowed range or the derivative with respect to $t$ is zero, i.e.,
\begin{equation}
\di{t}P_\mathrm{ss}=2\eta_2\left(\frac{s^2}{t^3q_{1b}}-\frac{t}{q_{1c}}\right)=0\,,
\end{equation}
which gives $q_{1c}=q_{1b}t^4/s^2$.
Together with the two constraints of \eqref{BCconstraints}, we have
$q_{2c}=\frac{t^2}{q_{1c}}=\frac{s^2}{t^2q_{1b}}=q_{2b}$.
In a similar way, we also have $q_{1c}=q_{1b}$ for the optimal solution due to the symmetry of the discrimination scheme for the two signal states. Thus, in order to optimize the joint probability of success, the success probabilities of Bob and Charlie must be equal, $q_{1b}=q_{1c}$ and $q_{2b}=q_{2c}$. Combining these with the constraints gives $s/t^2=\sqrt{\frac{q_{1\text{b}}q_{2\text{b}}}{q_{1\text{c}}q_{2\text{c}}}}=1$. Hence, $t=\sqrt{s}$. 
Upon eliminating two of the three parameters, $t(=\sqrt{s})$ and $q_{1c}(=q_{1b})$, $P_\mathrm{ss}$ becomes the function of the single parameter $q_{1b}$,
\begin{equation}\label{eq:SuccessProbq1b}
P_\mathrm{ss}=\eta_1(1-q_{1b})^2+\eta_2\left(1-\frac{s}{q_{1b}}\right)^2\,.
\end{equation}
For optimum, the derivative with respect to $q_{1b}$ must vanish, $\di{q_{1b}}P_\mathrm{ss}=0$, yielding a quartic equation,
\begin{equation}\label{eq:equationq1b}
\frac{\eta_1}{\eta_2}q_{1b}^3(1-q_{1b})-(q_{1b}-s)s=0\,.
\end{equation}
This equation has four solutions. The physical solutions must be real and in the range $s \leq q_{1b} \leq1$, depending on the value of $\eta_1,\eta_2$ and $s$. We first illustrate, as an example, the case for equal priors, $\eta_1=\eta_2$. For this case, the quartic equation can be solved analytically. Later, we extend the solution to arbitrary priors.

\begin{figure}[ht]
\centerline{\setlength{\unitlength}{1pt}
\begin{picture}(240,166)(0,0)
\put(5,70){\includegraphics[width=110pt]{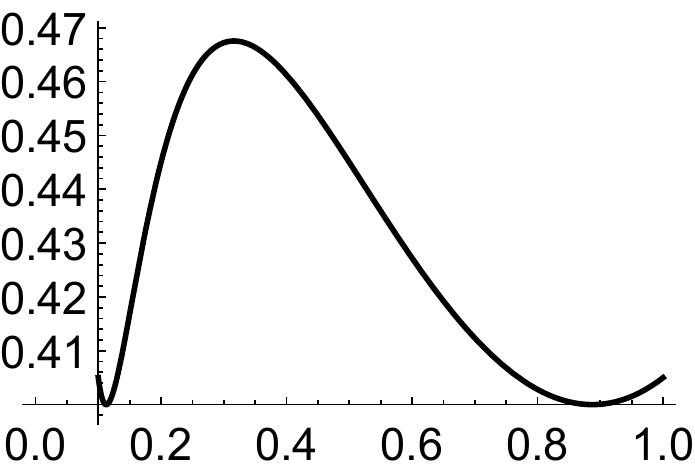}}
\put(120,70){\includegraphics[width=110pt]{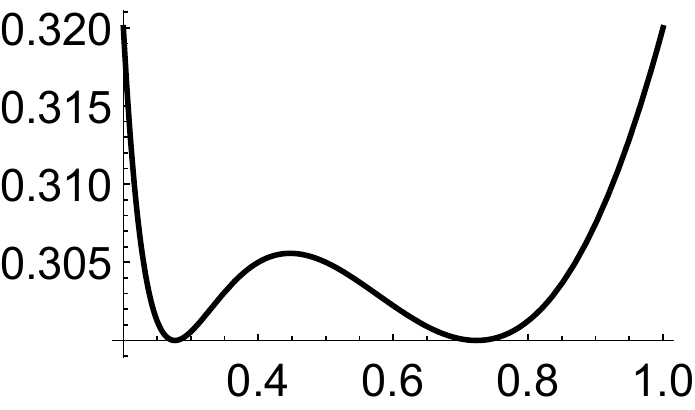}}
\put(5,-5){\includegraphics[width=110pt]{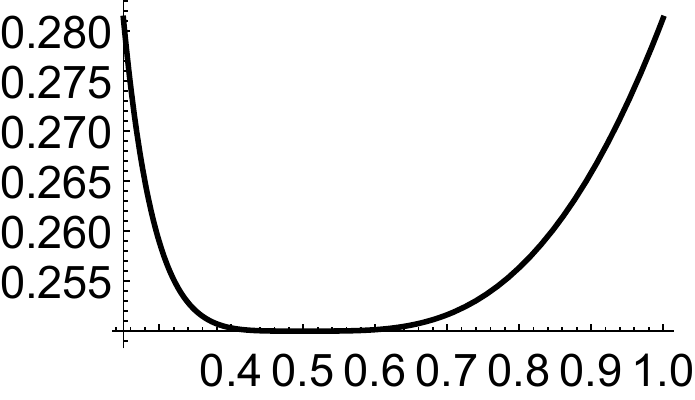}}
\put(120,-5){\includegraphics[width=110pt]{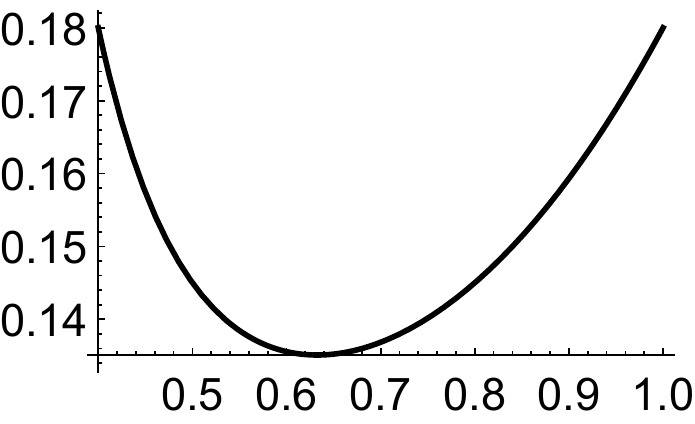}}
\put(80,120){\makebox(0,0){$s=0.1$}}
\put(180,120){\makebox(0,0){$s=0.2$}}
\put(60,40){\makebox(0,0){$s=0.25$}}
\put(180,40){\makebox(0,0){$s=0.4$}}
\put(114,154){\makebox(0,0){Joint success probabilty $P_\mathrm{ss}$, Eq. \eqref{eq:SuccessProbq1b}, vs. $q_{1b}$}}
\end{picture}}
\caption{The joint probability of success $P_\mathrm{ss}$ of \eqref{eq:SuccessProbq1b} as a function of $q_{1b}$ for $s=0.1$, 0.2, 0.25 and 0.4, and $\eta_1=\eta_2$. For each subfigure, the physical range of $s\leq q_{1b}\leq1$ is plotted. For $s\leq1/4$, the function has a local maximum at $q_{1b}=\sqrt{s}$ and two global minima at $q_{1b}=1/2(1\pm\sqrt{1-4s})$. At $s=s_{c}$ the local maximum is equal to the value on the boundary. The optimum of $P_\mathrm{ss}$ is at the local maximum for $s\leq s_{c}$, and on the boundary for $s>s_{c}$. For $s\geq1/4$ there is only one extremum which is a global minimum.}
\label{fig:PssUSDvsp}
\end{figure}
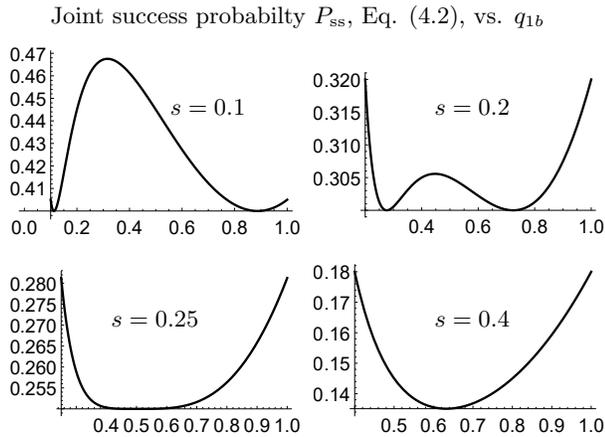

For equal priors, the four solutions of the equation are $\{\pm\sqrt{s}, 1/2(1\pm\sqrt{1-4s})\}$. For $s<1/4$, there are three physical solutions: $q_{1b}=\{\sqrt{s}, 1/2(1\pm\sqrt{1-4s})\}$. For $s\geq1/4$, there is only one physical solution at $q_{1b}=\sqrt{s}$. The solutions $q_{1b}=1/2(1\pm\sqrt{1-4s})$, if exist, always give the location of the minima of $P_\mathrm{ss}$. Thus, the maximum of $P_\mathrm{ss}$ must either be its value at the extremal point of $q_{1b}=\sqrt{s}$, or its value at the boundary solutions $q_{1b}=s$ or $q_{1b}=1$, shown in Fig.~\ref{fig:PssUSDvsp}.

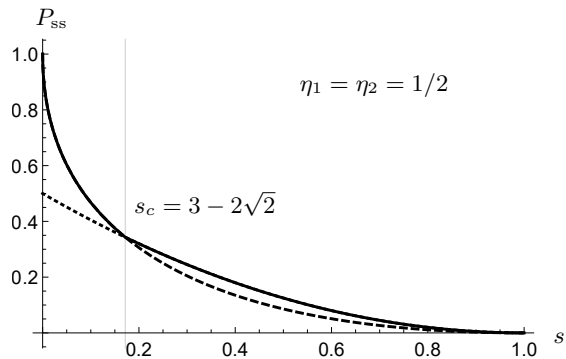
\begin{figure}[ht]
\centerline{\setlength{\unitlength}{1pt}
\begin{picture}(230,140)(0,0)
\put(13,-5){\includegraphics[scale=0.55]{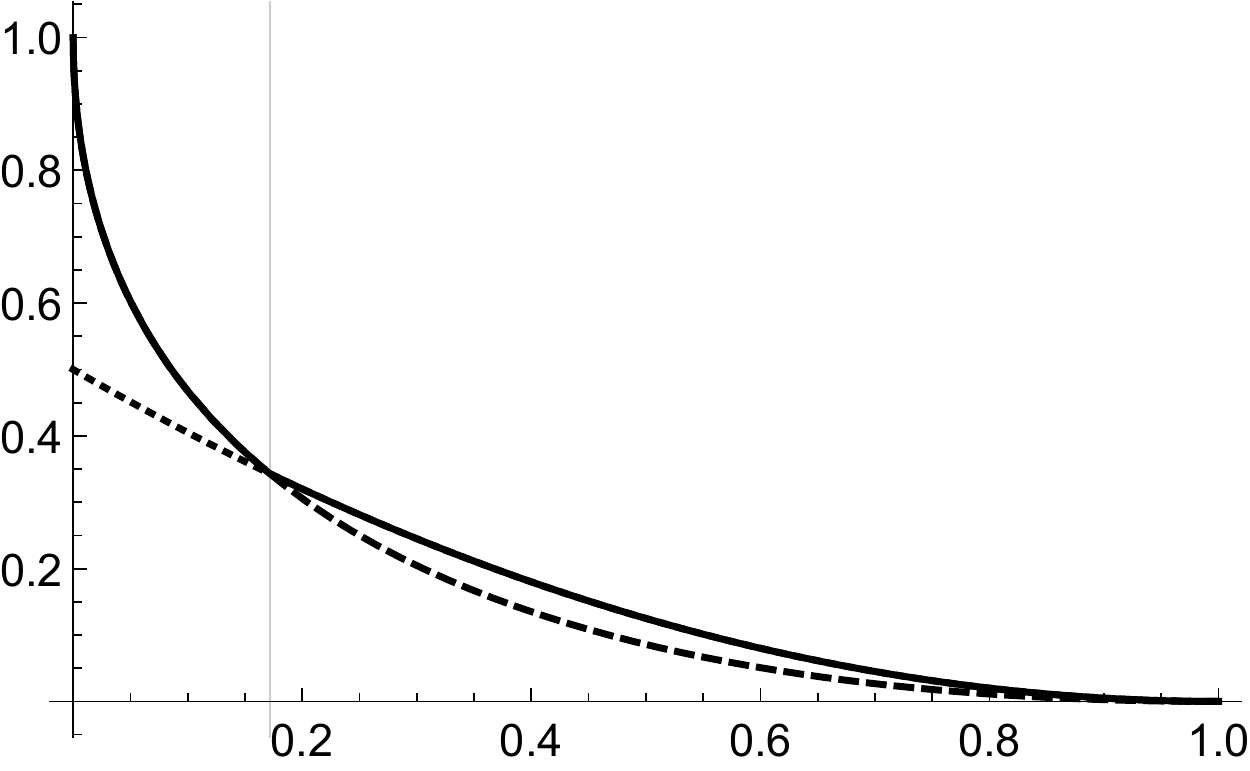}}
\put(220,4){\makebox(0,0){$s$}}
\put(28,125){\makebox(0,0){$P_\mathrm{ss}$}}
\put(86,55){\makebox(0,0){$s_c=3-2\sqrt{2}$}}
\put(150,100){\makebox(0,0){$\eta_1=\eta_2=1/2$}}
\end{picture}}
\caption{Solid line: the optimal joint success probability $P_\mathrm{ss}$, Eq. \eqref{pssmax}, vs. $s=\expectn{\psi_1|\psi_2}$ for equal priors. Dotted line: the boundary value solutions. Dashed line: the value of the function at $q_{1b}=\sqrt{s}$; the critical value of $s_c=3-2\sqrt{2}$ is the value at which these two curves intersect.}\label{fig:PssUSDequalP}
\end{figure}

Evaluating the joint probability of success at these values, we have $P_\mathrm{ss}(q_{1b}{=}\sqrt{s})=(1-\sqrt{s})^2$ at the local extremum, and $P_\mathrm{ss}(q_{1b}{=}s,1)=\frac{1}{2}(1-s)^2$ at the boundary. The value at the boundary is larger than the local maximum when $s > s_{c}=3-2\sqrt{2}\approx0.1716$. Thus,
\begin{eqnarray}
(P_\mathrm{ss})_\mathrm{max}=\left\{\begin{array}{l}(1-\sqrt{s})^2\;\; \mathrm{if}\;\;s\leq3-2\sqrt{2} \\ \frac{1}{2}(1-s)^2\;\;\;\mathrm{if}\;\;s>3-2\sqrt{2}\end{array}\right. .
\label{pssmax}
\end{eqnarray}
The dependence of the optimal joint probability of success on the overlap of the states $s$ is illustrated in  Fig.~\ref{fig:PssUSDequalP}.

\begin{figure}[ht]
\centerline{\setlength{\unitlength}{1pt}
\begin{picture}(240,136)(0,0)
\put(120,124){\makebox(0,0){$P_\mathrm{ss}$ of \eqref{eq:SuccessProbq1b} and the constraint $q_{1b}q_{2b}=s$}}
\footnotesize
\put(5,0){\includegraphics[width=110pt, height=110pt]{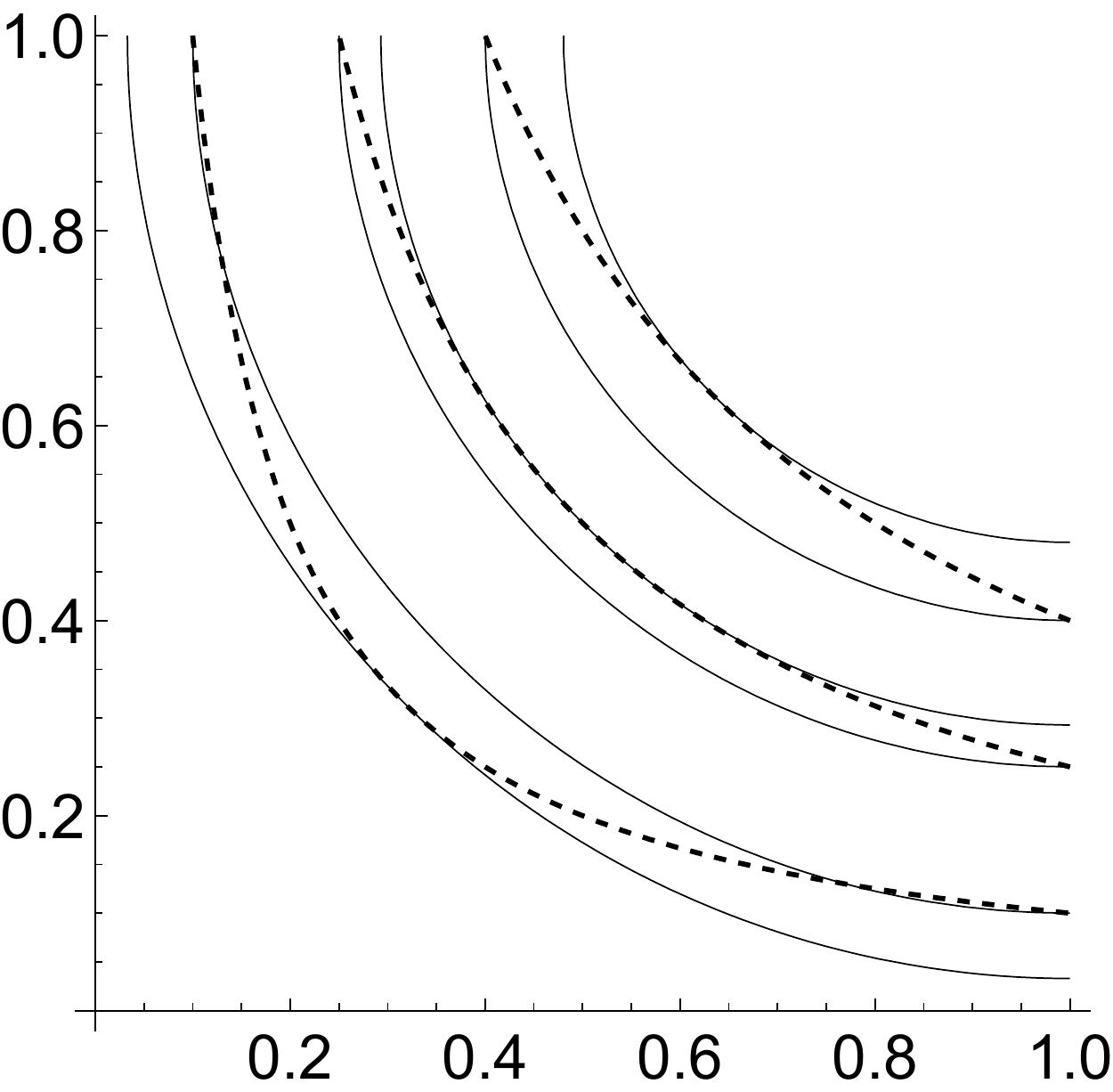}}
\put(130,0){\includegraphics[width=110pt, height=110pt]{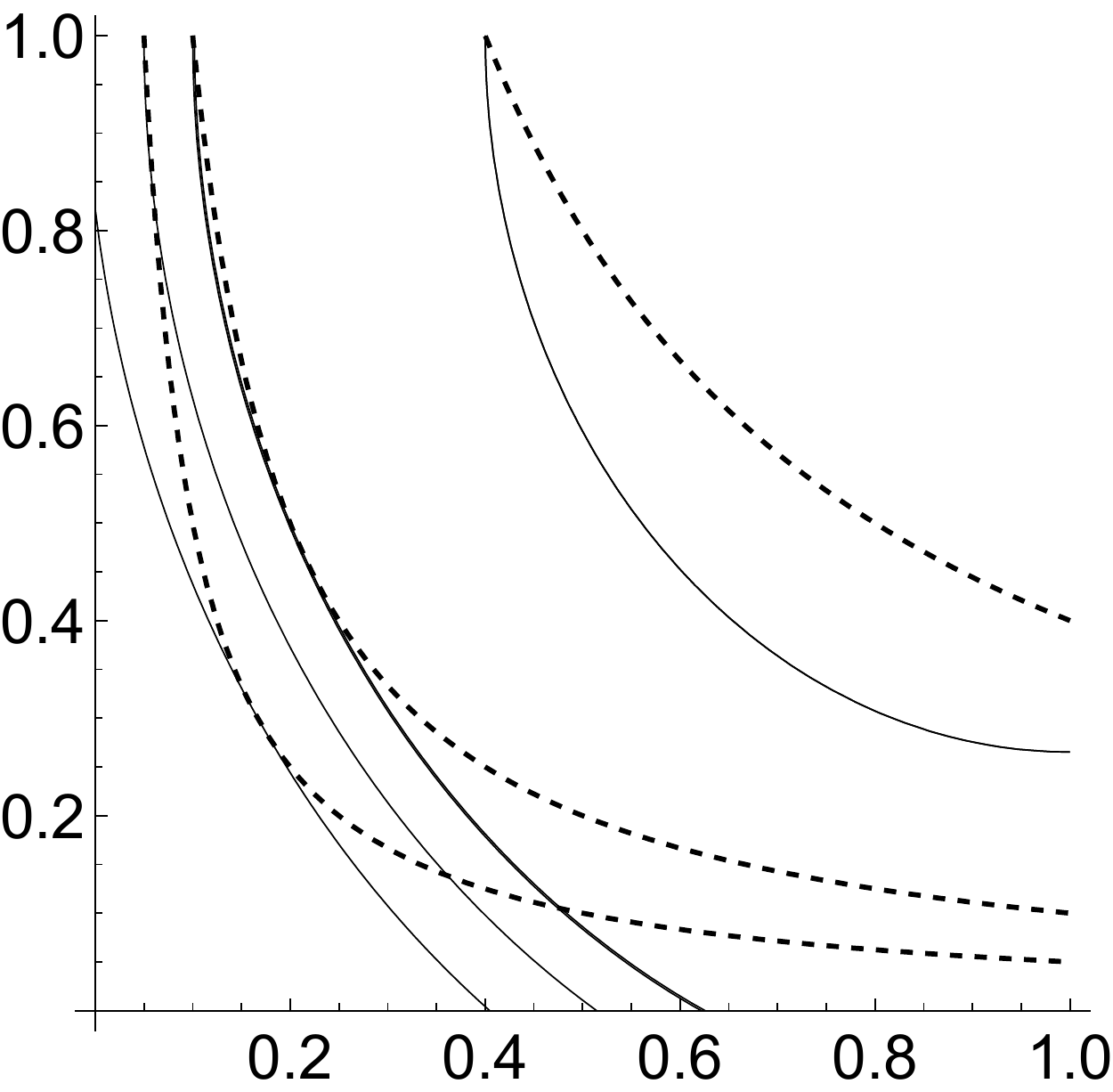}}
\put(2,56){\makebox(0,0){$q_{2b}$}}
\put(125,56){\makebox(0,0){$q_{2b}$}}
\put(62,-4){\makebox(0,0){$q_{1b}$}}
\put(188,-4){\makebox(0,0){$q_{1b}$}}
\put(88,100){\makebox(0,0){$\eta_1=\frac{1}{2}$}}
\put(208,100){\makebox(0,0){$\eta_1=\frac{3}{5}$}}
\put(36,36){\scriptsize\makebox(0,0){$0.1$}}
\put(66,56){\scriptsize\makebox(0,0){$0.25$}}
\put(84,72){\scriptsize\makebox(0,0){$0.4$}}
\put(150,30){\scriptsize\makebox(0,0){$0.05$}}
\put(172,45){\scriptsize\makebox(0,0){$0.1$}}
\put(202,78){\scriptsize\makebox(0,0){$0.4$}}
\end{picture}}
\caption{Contour plots of the joint probability of success $P_\mathrm{ss}$ of \eqref{eq:SuccessProbq1b} and the constraint $q_{1b}q_{2b}=s$ as functions of $q_{1b}$ and $q_{2b}$. Solid lines: contours of $P_\mathrm{ss}$. Dashed lines: plots of $q_{1b}q_{2b}=s$. The values of $s$ are used to label the lines.  For $\eta_1=\eta_2$ (left panel), contours of the joint probability of success $P_\mathrm{ss}$ are quarter segments of circles, symmetric under reflection about the line $q_{1b}=q_{2b}$; for $\eta_1\neq\eta_2$ (right panel), contours of the joint probability of success $P_\mathrm{ss}$ are segments of ellipses.}\label{fig:ConstraintContour}
\end{figure}
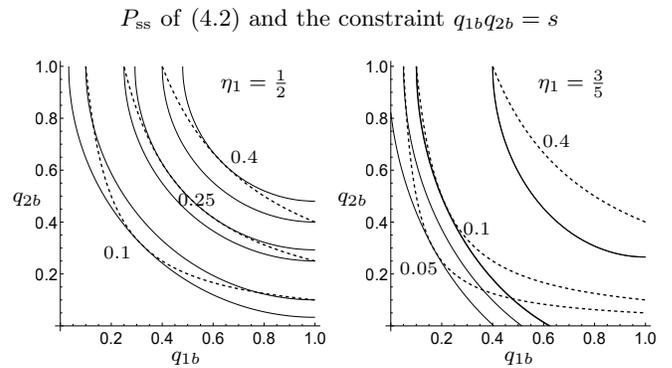

For general priors, $\eta_1 \neq \eta_2$, the optimal value of $P_\mathrm{ss}$ must also be either one of the physical solutions of \eqref{eq:equationq1b} in the interval $s < q_{1b} < 1$, or its value on the boundary, $q_{1b} = s$ or $q_{1b} = 1$. The two boundary solutions, however, are not the same as in the case of equal priors. If $\eta_1>\eta_2$, the value at the lower boundary, $q_{1b}=s$, is larger than the value at the upper boundary $q_{1b}=1$, and vice versa. The larger boundary value solution is given by 
\begin{equation}
P_\mathrm{ss}^{b}=\eta_{max}(1-s)^2\,,
\label{Pssboundary}
\end{equation}
where $\eta_{max} = \max\{\eta_1,\eta_2\}$. For every set of priors, there is a critical value of $s$ for which the boundary value solution of $P_\mathrm{ss}$ is the same as its value at the local maximum between $s\leq q_{1b}< 1$. This switching of the optimal value between the local maximum and the boundary values can be understood by the relation between $P_\mathrm{ss}$ and the constraint, shown in Fig.~\ref{fig:ConstraintContour}.

\begin{figure}[ht]
\centerline{\setlength{\unitlength}{1pt}
\begin{picture}(220,136)(0,0)
\put(0,-5){\includegraphics[width = 200pt]{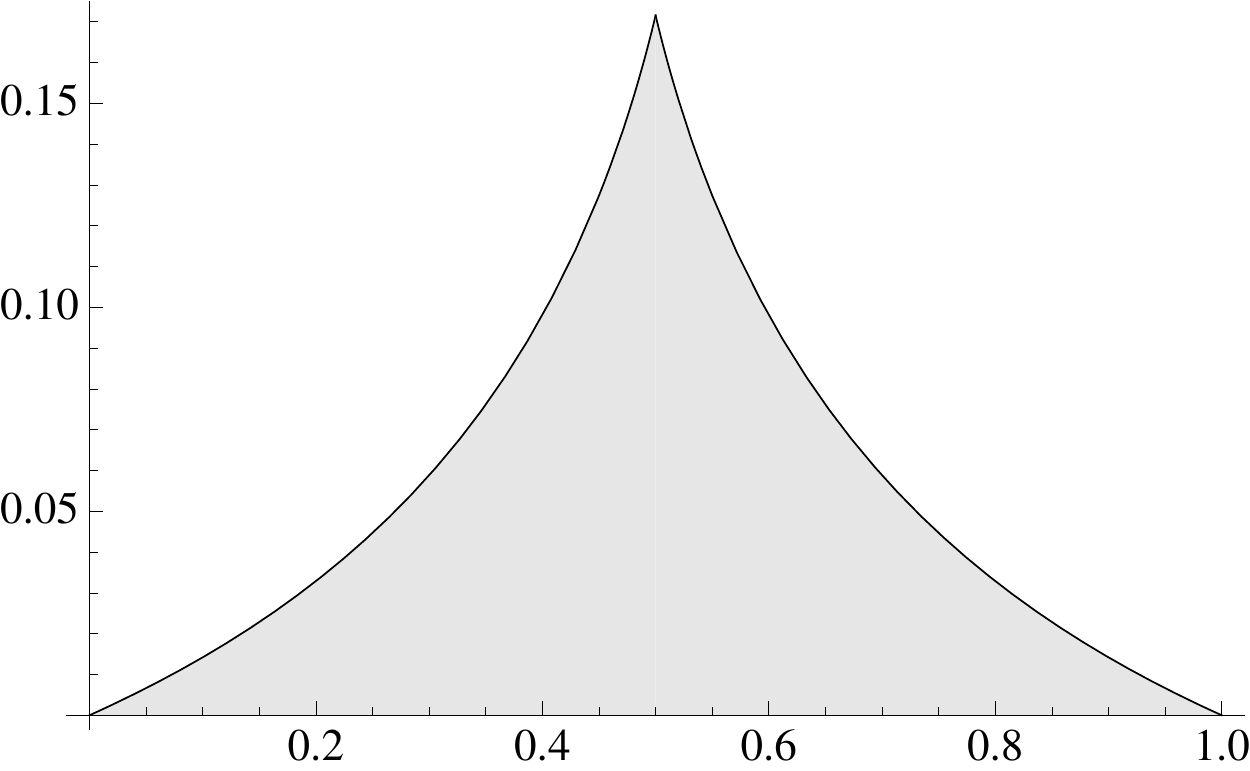}}
\put(210,0){\makebox(0,0){$\eta_1$}}
\put(15,124){\makebox(0,0){$s_c$}}
\end{picture}}
\caption{The critical value for the overlap of states $s_c$ vs. the prior probability $\eta_1$.  The shaded area indicates the parameter regime where the nontrivial solution for the local maximum of $P_\mathrm{ss}$ is larger than the boundary solutions.}\label{fig:PssUSDsCritical}
\end{figure}

At the critical value of $s$, $s_c$,  $P_\mathrm{ss} = P_{ss}^{b}$. For $s < (>) s_c$, we have $P_\mathrm{ss} > (<) P_{ss}^{b}$. The dependence of $s_c$ on the prior probability $\eta_1$ is shown in Fig.~\ref{fig:PssUSDsCritical}.

The critical value $s_c=3-2\sqrt{2}\approx0.1716$ for equal priors. $s_c$ decreases as the priors become more biased. The parameter region where the local maximum of \eqref{eq:SuccessProbq1b} is the optimal value for the joint probability of success, shown by the shaded region in Fig.~\ref{fig:PssUSDsCritical}, is quite small compared to the entire parameter regime of $s$ and $\{\eta_1,\eta_2\}$, which is given by the unit square $0 \leq s,\eta_{1} \leq 1$. Thus, for most of the range of $s$ and the priors, $P_\mathrm{ss}$ is optimized at the boundary, $q_{1(2)}^{boundary} = s$ and $q_{2(1)}^{boundary} = 1$ for $\eta_{1} > (<) 1/2$. In this case, both Bob and Charlie fail to detect state $\ket{\psi_2}$ (or $\ket{\psi_1}$) at all time but they optimize their set up such that state $\ket{\psi_1}$ (or $\ket{\psi_2}$) is successfully identified with probability $1-s$.

Although the joint probability of success can be optimized by the boundary solutions, the information that Bob and Charlie share with each other and with Alice is useless for communication as they only get a string of identical bits, after discarding the inconclusive outcomes. For example, in the case of $\eta_1>\eta_2$, they share a string of 0's which carries no useful information. However, in the next section we discuss a measurement scheme that salvages the boundary solution and makes it useful for communication purposes.

\section{The Flip-Flop Measurement}
\label{AppC}

In Sec. II and also at the end of the previous section, we found that for a large range of the overlap parameter, $s$, the measurement that optimizes the joint probability of success is the one which unambiguously identifies one of the states and misses the other completely. Performing this measurement cannot transmit information that is useful for quantum communication. 

It was noticed, however, in the case of two-party communication between Alice and Bob (e.g. in the B92 cryptography protocol \cite{Bennett92}) that the von Neumann setup can be used to generate a random key.  In this case, Bob randomly choses between the two von Neumann setups, one that projects on $\{|\psi_{1}\rangle, |\psi_{1}^{\perp}\rangle\}$ and the other that projects on $\{|\psi_{2}\rangle, |\psi_{2}^{\perp}\rangle\}$.  For the first setup, $P_{0}^{(1)}=|\psi_{1}\rangle \langle\psi_{1}|$ is the inconclusive detector since a click in this detector may originate from either of the input states, and $I - P_{0}^{(1)}=|\psi_{1}^{\perp}\rangle \langle\psi_{1}^{\perp}|$  is the one that unambiguously identifies the input as $|\psi_{2}\rangle$ since it never clicks for $|\psi_{1}\rangle$. The action of the second setup can be obtained by interchanging the indices 1 and 2. 

In the flip-flop measurement Bob randomly chooses between the two setups. With probability $c$ he chooses the first setup and with probability $1 - c$ he choses the second setup. What this means is that Bob effectively flip-flops between the two von Neumann setups. The failure probability, averaged over the flipping rate, is  
\begin{eqnarray}
q_{1} &=& c \langle\psi_{1}|P_{0}^{(1)}|\psi_{1}\rangle  + (1-c) \langle\psi_{1}|P_{0}^{(2)}|\psi_{1}\rangle  \nonumber \\
&=& c + (1-c) s^{2} 
\label{cq1}
\end{eqnarray} 
for the first state, and  
\begin{eqnarray}
q_{2} &=& c \langle\psi_{2}|P_{0}^{(1)}|\psi_{2}\rangle  + (1- c)\langle\psi_{2}|P_{0}^{(2)}|\psi_{2}\rangle \nonumber \\ 
&=& 1- c + cs^{2}
\label{cq2}
\end{eqnarray} 
for the second. Clearly, $q_{1}q_{2} = s^{2} + c(1-c)(1-s^{2})^{2}  \ge s^{2}$, so this procedure is not optimal unless $c=0$ or $c=1$. 

The success probability averaged over the flipping rate is $p_{1} = 1 - q_{1} = (1-c)(1-s^2)$ for the first state, and $p_{2} =  1 - q_{2} = c (1-s^2)$ for the second. Thus, the average success probability for the flip-flop measurement is
\begin{equation}
P_{\text{succ}} = \eta_{1} p_{1} + \eta_{2} p_{2} = [\eta_{1} (1-c)+ \eta_{2} c](1 - s^{2})\,.
\label{cPs}
\end{equation}
The average probability of failure, $Q$, is $Q=1-P_{\text{succ}}$. 

$P_{\text{succ}}$ is a linear function of the flipping rate $c$, so the function is either monotonically increasing, monotonically decreasing or constant. If $\eta_{1}=\eta_{2}=1/2$, the function is constant. Otherwise, the maximum is on one of the boundaries of the $0 \leq c \leq 1$ interval. $P_{\text{succ}}$ is maximum for $c=0$ when $\eta_{1} >\eta_{2}$ and for $c=1$ when $\eta_{1}<\eta_{2}$. The strategy that maximizes the success probability is to always bet on the state with the larger prior probability. 

The flip-flop measurement for $0 < c < 1$ has a lower success probability than the optimal boundary solution. However, it can generate a bit string that contains both 0's and 1's, not just one of them. In this respect, one particular choice of $c$ stands out. For $c=\eta_{1}$, the two terms on the r.h.s. of \eqref{cPs} become equal, yielding $P_{\text{succ}} = 2 \eta_{1} \eta_{2} (1-s^{2})$. In this case, the flip-flop measurement generates a random string of 0's and 1's where the occurrence probability of the 0's is equal to that of the 1's, a very desirable feature for QKD applications.   

After the discussion of the flip-flop measurement on the example of two-party communication, we now extend these considerations to the sequential UD scheme. In the sequential version of the flip-flop measurement both Bob Charlie choose randomly between the two boundary setups. For simplicity, we assume that their flipping rates are equal. Independently, each with probability $c$ chooses the first  setup and with probability $1 - c$ the second setup. Their failure probabilities, averaged over $c$, are    

\begin{eqnarray}
	q_{1\text{b}} &=&  c  +(1 -c) s\,, \label{Sequential FFM q1} \\
	q_{1\text{c}} &=&  c  +(1 -c) s\,,  \\
	q_{2\text{b}} &=&  c s  +(1 -c) \,, \\
	q_{2\text{c}} &=&  c s  +(1 -c) \,.
	\label{Sequential FFM q}
\end{eqnarray}
The corresponding success probabilities averaged over the flipping rate are
\begin{eqnarray}
	p_{1\text{b}} &=&  (1-c) (1-s )\,,  \\
	p_{1\text{c}} &=&  (1-c) (1-s)\,,   \\
	p_{2\text{b}} &=&  c (1-s)\,, \\
	p_{2\text{c}} &=&  c (1-s) \,.
	\label{Sequential FFM p}
\end{eqnarray}
The average joint probability of success for the flip-flop measurement is, thus, given by 
\begin{eqnarray}
P_{\textrm{ss}}^{\textrm{(f)}} =[ (1-c)^{2}\eta_{1} + c^{2}\eta_{2}](1-s)^{2} .
\label{FF1}
\end{eqnarray}

This is a simple quadratic function of the flipping rate, $c$, reaching its maximum at $c=0$ if $\eta_{1} > \eta_{2}$ and at $c=1$ if 
$\eta_{2} > \eta_{1}$. Inside the interval $0 < c < 1$, it reaches its minimum when $c=\eta_{1}$, 
\begin{equation}
P_{\textrm{ss,min}}^{\textrm{(f)}} = \eta_{1} \eta_{2} (1-s)^{2} \,.
\label{Pfssmin}
\end{equation}
Clearly, this is the worst strategy for unambiguous identification of the states prepared by Alice. However, what is worst for one thing is best for another. This strategy will generate an unbiased bit string of 0's and 1's, so this is the best strategy for application in QKD or, in general, quantum communication schemes. 
 
\section{Mutual Information}

\subsection{Unambiguous communication channel}

One disadvantage of the optimal solution occurring on the boundary is that it does not lead to a quantum communication protocol between Alice and Bob (and Charlie). Bob effectively ignores one of the states, setting the probability of successfully detecting that state to 0. If Bob restricts himself to only keeping a result when it is conclusive, he will end up with a string of identical bits. This is, of course, useless for establishing a secret key with Alice and for communication purposes, in general. 

The amount of information transmitted is quantified by the mutual information. We adopt the common convention of denoting the message of the sender by $X$ and the message the receiver decoded by $Y$. The mutual information of the communication channel is defined as
\begin{equation}
	I\left( A:B \right) = H(X) - H(X|Y).
	\label{Mutual information}
\end{equation}
$H(X)=H(\eta_1)\equiv=-\eta_1\log_2\eta_1-(1-\eta_1)\log_2(1-\eta_1)$ denotes the Shannon entropy of the sender's binary information and $H(X|Y)$ denotes the conditional Shannon entropy \cite{Cover2006}. For a general three-element POVM $\{\Pi_1,\Pi_2,\Pi_0\}$, the mutual information is given \cite{Peres2002}, by
\begin{equation}
	I(A:B) = H(\eta_{1}) - \sum_{j=0}^2P\left( \Pi_{j} \right)H\left(X|\Pi_{j}\right),	\label{ud mutual information}
\end{equation}
where $P(\Pi_j)$ denotes the probability of having measurement outcome $\Pi_j$.
If Bob gets a click in either the $\Pi_{1}$ or $\Pi_{2}$ detectors, he has no uncertainty as to what state Alice sent, therefore $H(X|\Pi_{1})=H(X|\Pi_{2}) = 0$. If $\left\{ q_{1},q_{2} \right\}$ represent the failure probabilities when Bob attempts to detect states $\left\{ \ket{\psi_{1}},\ket{\psi_{2}} \right\}$, then $P\left( \Pi_{0} \right) = \eta_{1}q_{1} + \eta_{2}q_{2} \equiv Q$ and $H\left( X|\Pi_{0} \right) = H\left( \frac{\eta_{1}q_{1}}{Q} \right)$. Plugging these values in to \eqref{ud mutual information}, we have
\begin{equation}
	I\left( A:B \right) = H\left( \eta_{1} \right) - QH\left( \frac{\eta_{1}q_{1}}{Q} \right).
	\label{UDMutualReduced}
\end{equation}
The mutual information is maximized when $QH\left( \frac{\eta_{1}q_{1}}{Q} \right)$ is minimized. 

The calculation suggests that information is maximally transmitted when Bob detects only one of the two incoming states, which is a counterintuitive result. In order to resolve this quandary, one should realize that this formulation treats all three detection outcomes by Bob, $\left\{ \Pi_{1},\Pi_{2},\Pi_{0} \right\}$, on equal footing, unambiguous discrimination not playing any distinguished role. If the outcome is $\Pi_{0}$, Bob guesses which state Alice sent him based on which state was more likely to have failed. He will make some errors, but will still obtain some information. It is clear that Bob succeeds in this strategy the most when the $\Pi_{0}$ channel produces the least uncertainty, which is the result calculated. In this formalism, Bob treats the inconclusive and the conclusive outcomes in the same way, and does not share with Alice when his outcome is inconclusive to discard those results. This approach is closer to the minimum error state discrimination strategy as errors are permitted but is not quite suitable for the unambiguous discrimination strategy.

The mutual information for a truly unambiguous channel has to take into account that only error-free messages are kept. The outcome is conclusive with probability $P_s$, and inconclusive with probability $Q$. Hence, after discarding the inconclusive outcomes, the mutual information, conditioned on success, is given by
\begin{equation}\label{eq:MutualInfoUSD2}
I_\mathrm{USD}(A:B)=P_s\Big[H(X_c)-\underbrace{H(X_c|Y_c)}_{=0}\Big]=P_sH(X_c)\,,
\end{equation}
where $X_c$ and $Y_c$ denotes the messages of the sender and the receiver for conclusive outcomes, respectively. $H(X_c|Y_c)=0$ because there is no uncertainty among conclusive outcomes (i.e., $X_c=Y_c$). 
The prior probability for Alice's message $X_c$ is given by the confidence probabilities $\{C_{s,1},1-C_{s,1}\}$ for states $\{\ket{\psi_1},\ket{\psi_2}\}$, where
\begin{equation}
C_{s,1}=\frac{\eta_1(1-q_1)}{P_s}\,.
\end{equation}
Hence, the correct expression of the conditional mutual information for USD is
\begin{equation}
I_\mathrm{USD}(A:B)=P_sH(X_c)=P_sH(C_{s,1})\,.\nonumber
\end{equation}
With this expression, it is clear that if Bob restricts himself to only gaining information from error-free results, the amount of information gained by the boundary solutions, i.e., when either $q_{1}$ or $q_{2}$ are set to $1$, is zero.

The fundamental difference between the mutual information $I(A{:}B)$ of Eq.~\eqref{ud mutual information} and $I_\mathrm{USD}(A{:}B)$ of Eq.~\eqref{eq:MutualInfoUSD2} comes from Bob sharing the classical information of whether his measurement outcome is conclusive. Upon having this classical information, the mutual information of this quantum communication channel is reduced to $I_\mathrm{USD}(A{:}B)$ even if we take into account all of the measurement outcomes including the inconclusive ones. Alice's Shannon entropy can be divided into the uncertainty coming from the conclusive outcomes and the uncertainty coming from the inconclusive ones, i.e., $H_\mathrm{USD}(X)=P_\mathrm{s}H(X_c)+QH(X_{inc})$. The conditional entropy, $H(X|Y)=P_s\cdot0+QH(X|\Pi_0)=QH(X_{inc})$. Thus, the mutual information given by the difference is $I_\mathrm{USD}(A:B)=P_\mathrm{s}H(X_c)$. This shows that, although Bob can obtain information from the inconclusive outcomes, this part of the information is shared publicly, including Alice or an eavesdropper, through classical communication and not through quantum communication. 

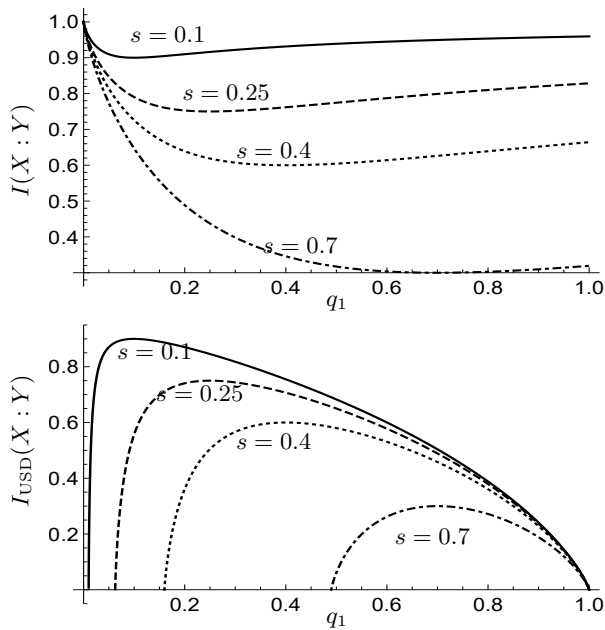
\begin{figure}[ht]
\centerline{\setlength{\unitlength}{1pt}
\begin{picture}(220,240)(0,0)
\put(5,120){\includegraphics[width=210pt, height=110pt]{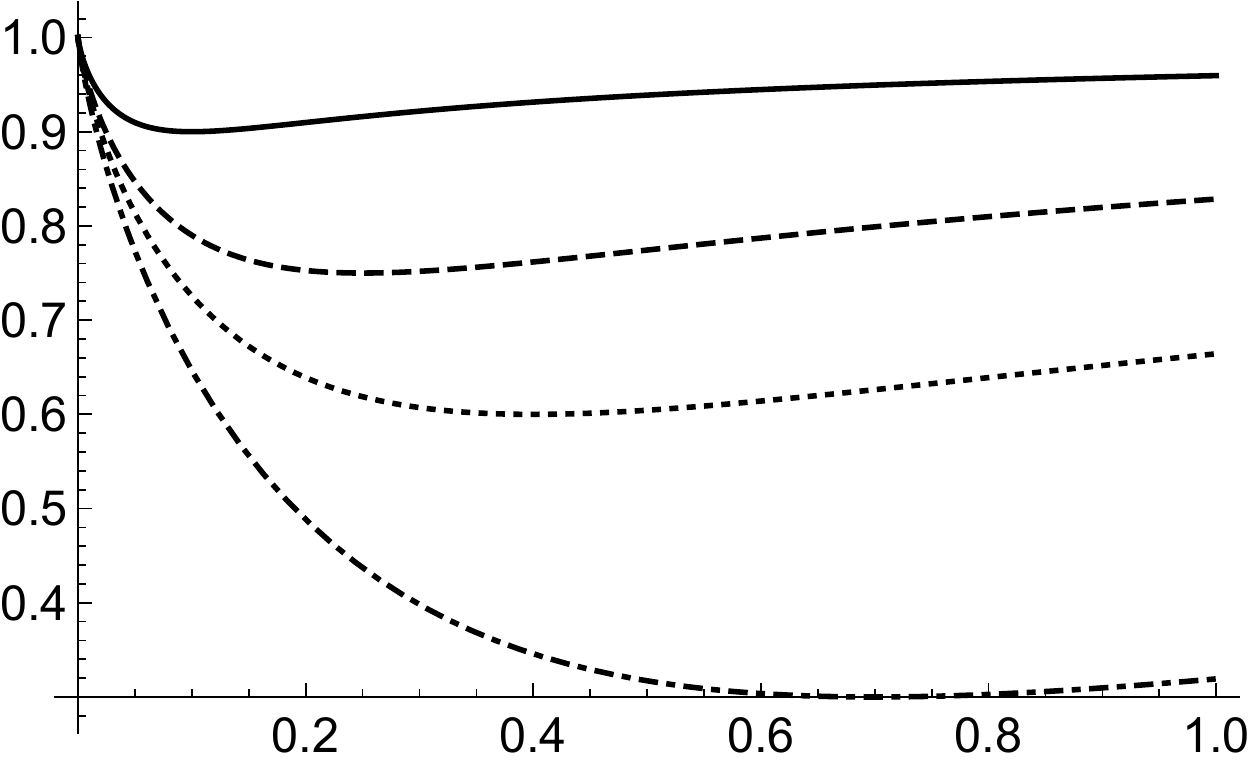}}
\put(5,0){\includegraphics[width=210pt, height=110pt]{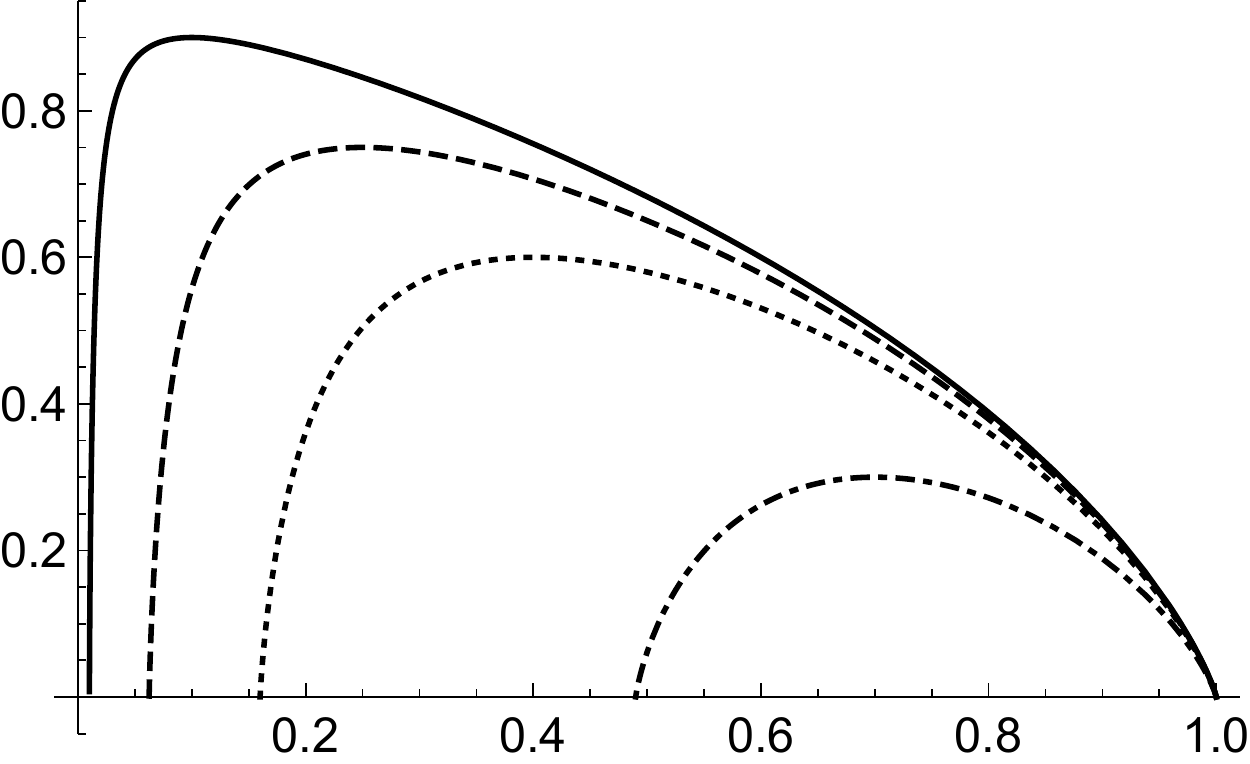}}
\put(114,-2){\makebox(0,0){$q_1$}}
\put(114,118){\makebox(0,0){$q_1$}}
\put(-5,176){\makebox(0,0){\rotatebox{90}{$I(X:Y)$}}}
\put(-5,66){\makebox(0,0){\rotatebox{90}{$I_\mathrm{USD}(X:Y)$}}}
\put(50,220){\makebox(0,0){$s=0.1$}}
\put(72,198){\makebox(0,0){$s=0.25$}}
\put(90,176){\makebox(0,0){$s=0.4$}}
\put(100,140){\makebox(0,0){$s=0.7$}}
\put(45,100){\makebox(0,0){$s=0.1$}}
\put(62,84){\makebox(0,0){$s=0.25$}}
\put(90,66){\makebox(0,0){$s=0.4$}}
\put(150,30){\makebox(0,0){$s=0.7$}}
\end{picture}}
\caption{Mutual information $I(A{:}B)$, Eq.~\eqref{ud mutual information} (upper), and $I_\mathrm{USD}(A{:}B)$, Eq.~\eqref{eq:MutualInfoUSD2} (lower), vs. $q_1$, for $\eta_1=\eta_2=1/2$. Solid line: $s=0.1$. Dotted line: $s=0.25$. Dashed line: $s=0.4$. Dot-dashed line: $s=0.7$. We choose $q_2=s^2/q_1$ for the optimal USD.}
\label{fig:HBUSDConShannon}
\end{figure}

$I(A:B)$ and $I_{USD}(A:B)$, Eq. \eqref{ud mutual information} and Eq. \eqref{eq:MutualInfoUSD2}, are compared in Fig. \ref{fig:HBUSDConShannon}.

It is important to notice that, for $\eta_1=\eta_2$, the maximum of $I_\mathrm{USD}(A{:}B)$ and the minimum of $I(A{:}B)$ occurs at the nontrivial solution for the optimization of the probability of success.

\subsection{Optimization of the mutual information}

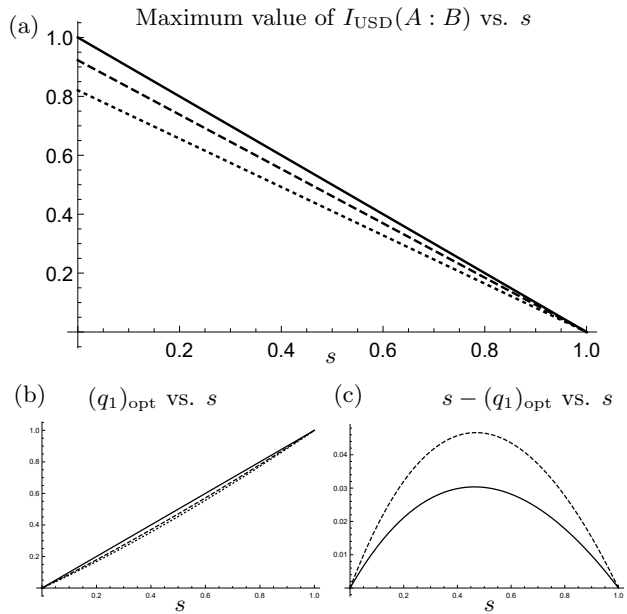
\begin{figure}[ht]
\centerline{\setlength{\unitlength}{1pt}
\begin{picture}(230,240)(0,0)
\put(6,90){\includegraphics[width = 210pt]{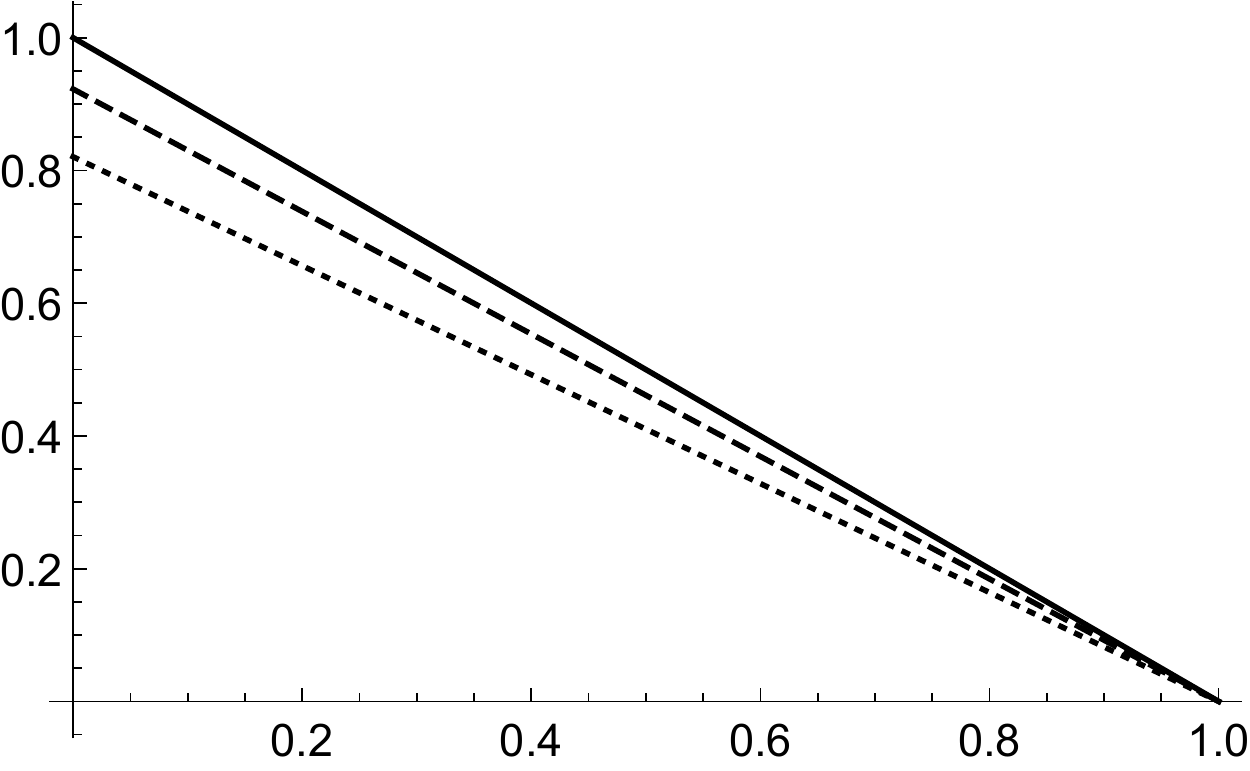}}
\put(0,0){\includegraphics[width = 110pt]{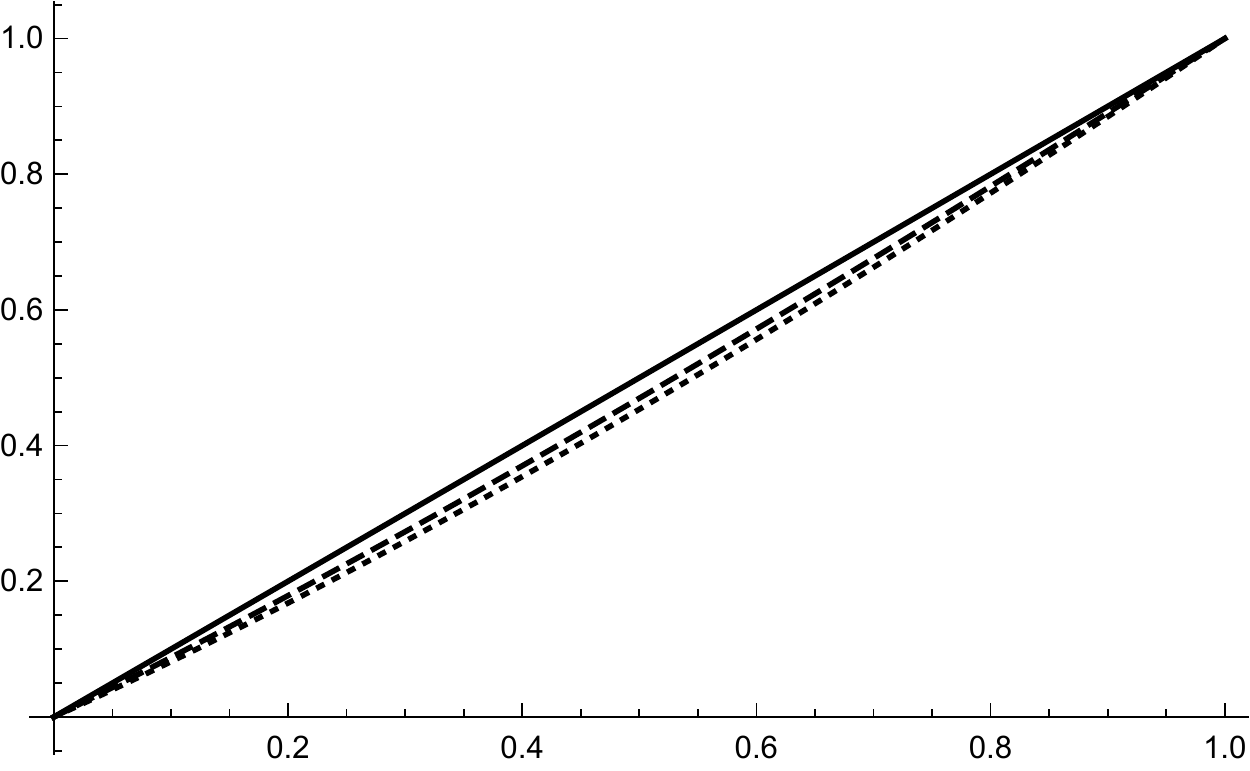}}
\put(115,0){\includegraphics[width = 110pt]{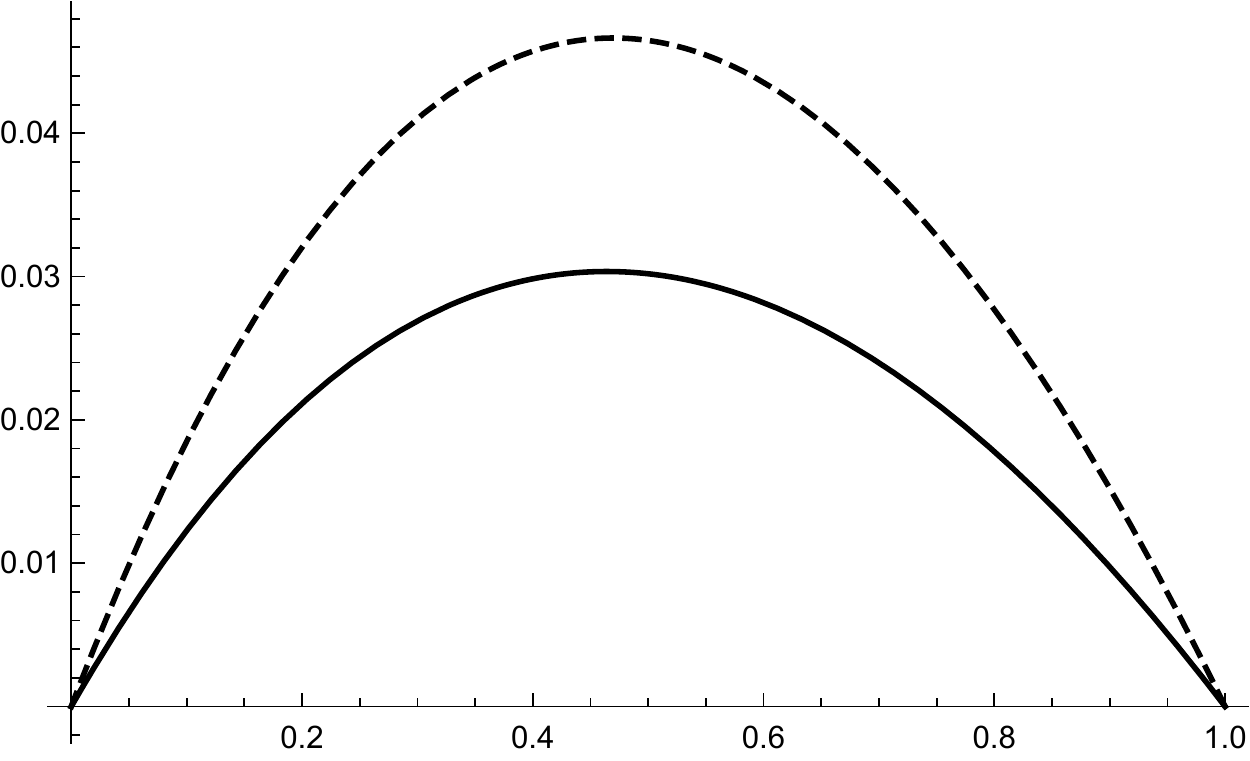}}
\put(114,92){\makebox(0,0){$s$}}
\put(57,-2){\makebox(0,0){$s$}}
\put(174,-2){\makebox(0,0){$s$}}
\put(115,220){\makebox(0,0){Maximum value of $I_\mathrm{USD}(A:B)$ vs. $s$}}
\put(22,74){\makebox(0,0)\footnotesize{$(q_1)_\mathrm{opt}$ vs. $s$}}
\put(156,74){\makebox(0,0)\footnotesize{$s-(q_1)_\mathrm{opt}$ vs. $s$}}
\put(-8,215){\makebox(0,0)\scriptsize{(a)}}
\put(-6,75){\makebox(0,0)\scriptsize{(b)}}
\put(115,75){\makebox(0,0)\scriptsize{(c)}}
\end{picture}}
\caption{(a) Upper bounds of the mutual information $I_\mathrm{USD}(A:B)$ for the unambiguous communication channel between Alice and Bob as a function of the overlap of states $s=\expectn{\psi_1|\psi_2}$. Solid line: $\eta_1=1/2$. Dashed line: $\eta_1=1/3$. Dotted line: $\eta_1=1/4$. These lines are visually indistinguishable from the plots of the approximate solution,  Eq.~\eqref{eq:MuInfoABAppro}, with $q_1=s$. (b) The plot of the values of $q_1$ that optimize $I_\mathrm{USD}(A:B)$ as a function of $s$. (c) The difference between the optimal value of $q_1$ and $q_1=s$.}
\label{fig:HBUSDMuInfoMax}
\end{figure}

Upon choosing $q_2=s^2/q_1$ for optimal USD, ensuring that no information is left in the post-measurement states,  $I_\mathrm{USD}(X:Y)$ becomes the function of the single parameter $q_1$. It is optimized when its derivative with respect to $q_1$ vanishes, i.e.,
\begin{equation}\label{eq:diInfoUSD}
\di{q_1}I_\mathrm{USD}=\eta_1\log_2\frac{\eta_1(1-q_1)}{P_s}-\eta_2\frac{s^2}{q_1^2}\log_2\frac{\eta_2(1-q_2)}{P_s}=0\,.
\end{equation}
This equation has a simple solution $q_1=q_2=s$ for the case of equal priors, which is same as the local maximum solution for the success probability $P_s$. Since the mutual information is a concave function within the physical range of the parameter $s^2\leq q_1\leq1$, the solution of \eqref{eq:diInfoUSD} maximizes the mutual information $I_\mathrm{USD}(A:B)$.
For $\eta_1=\eta_2=1/2$, $H(X_c)=1$ is maximized and $P_s$ is at its local extrema when $q_1=s\sqrt{\eta_2/\eta_1}=s$. Thus, $q_1=q_2=s$ must be the solution for equal priors and the optimal mutual information is $I_\mathrm{USD}(A:B)=1-s$. For the case of unequal priors $\eta_1\neq\eta_2$, however, we are not able to solve the equation analytically and have to rely on numerical methods, presented in Fig.~\ref{fig:HBUSDMuInfoMax}.

Figure~\ref{fig:HBUSDMuInfoMax}(a) displays $I_\mathrm{USD}(A:B)$ vs. $s$. For equal priors, the analytical upper bound is a linear function of $s$, $I_\mathrm{USD}(A:B)\leq1-s$. For $\eta_1\neq\eta_2$, its dependence on $s$ is almost linear. Fig.~\ref{fig:HBUSDMuInfoMax}(b) and (c) show that the value of $q_1$, that maximizes $I_\mathrm{USD}(A:B)$, depends only weakly on the priors. The difference between $(q_1)_\mathrm{opt}$ for arbitrary priors and $(q_1)_\mathrm{opt}=s$ for equal priors is largest at, and symmetric about, $s=1/2$. The dashed curve shows the approximate upper bound $I_\mathrm{USD}(A:B)$ for $q_1=s$,
\begin{equation}\label{eq:MuInfoABAppro}
I_\mathrm{USD}(A:B)(q_1=s)=(1-s)H(\eta_1)\,.
\end{equation}
We can conclude that for $\eta_{1}=\eta_{2}$, $I_\mathrm{USD}(A:B)$ is exactly optimized by the local maximum of the probability of success obtained at $q_1=q_2=s$. For $\eta_{1} \neq \eta_{2}$, the optimal value of $I_\mathrm{USD}(A:B)$ remains extremely close to the value given by $q_1=q_2=s$, i.e., to $I_\mathrm{USD}(A:B)=(1-s)H(\eta_1)$. 

\subsection{The sequential measurement scheme}

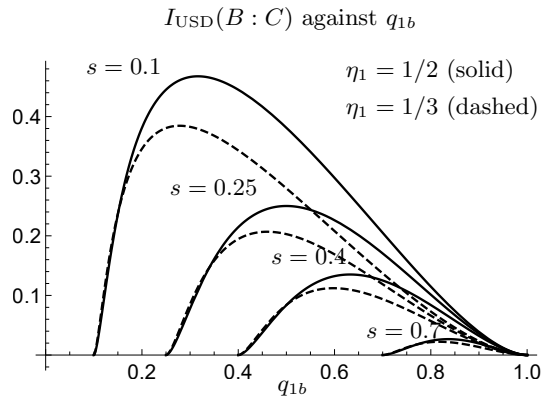
\begin{figure}[ht]
\centerline{\setlength{\unitlength}{1pt}
\begin{picture}(220,150)(0,0)
\put(0,0){\includegraphics[width = 200pt]{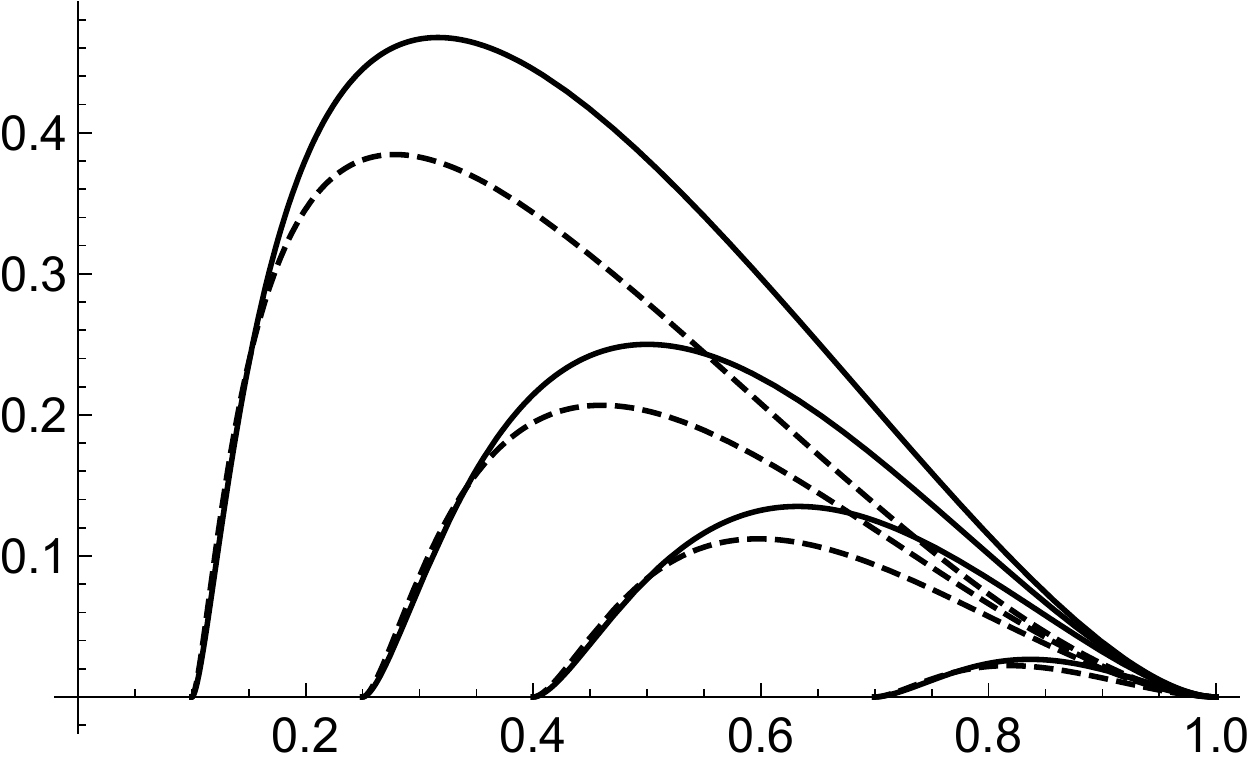}}
\put(42,120){\makebox(0,0){$s=0.1$}}
\put(76,74){\makebox(0,0){$s=0.25$}}
\put(112,48){\makebox(0,0){$s=0.4$}}
\put(148,20){\makebox(0,0){$s=0.7$}}
\put(106,-2){\makebox(0,0){$q_{1b}$}}
\put(105,138){\makebox(0,0){$I_\mathrm{USD}(B:C)$ against $q_{1b}$}}
\put(126,116){\makebox(0,0)\scriptsize{$\eta_1=1/2$ (solid)}}
\put(126,102){\makebox(0,0)\scriptsize{$\eta_1=1/3$ (dashed)}}
\end{picture}}
\caption{Mutual information between Bob and Charlie for the sequential USD channel, $I_\mathrm{USD}(B:C)$ from  Eq.~\eqref{eq:MuInfoBCSymmetric}, as a function of $q_{1b}$ for different values of $s$.}
\label{fig:BCMutualInfoEqual}
\end{figure}

For the sequential measurement scheme discussed in Sec. IV, Bob's probability of success to correctly identify Alice's message is $P_\mathrm{sb}=\eta_1(1-q_{1b})+\eta_2(1-q_{2b})$ and the probability for Charlie to correctly identify Alice's message is $P_\mathrm{sc}=\eta_1(1-q_{1c})+\eta_2(1-q_{2c})$. Keeping only conclusive outcomes, the mutual information for the communication channel between Alice and Bob and, respectively, between Alice and Charlie are,
\begin{eqnarray}
	I_\mathrm{USD}(A:B) &=& P_{\text{sb}}H\left(\frac{\eta_{1}(1-q_{1\text{b}})}{P_{\text{sb}}} \right), \\
	I_\mathrm{USD}(A:C) &=& P_{\text{sc}}H\left( \frac{\eta_{1}(1-q_{1\text{c}})}{P_{\text{sc}}} \right).
	\label{SequentialMuInforABC}
\end{eqnarray}
The mutual information between Bob and Charlie takes into account the events where both of them have successfully identified Alice's message, so that they share an identical string of bits. Their joint probability of success, $P_\mathrm{ss}=\eta_{1}p_{1\text{b}}p_{1\text{c}} + \eta_{2}p_{2\text{b}}p_{2\text{c}}=\eta_1(1-q_{1b})(1-q_{1c})+\eta_2(1-q_{2b})(1-q_{2c})$, was given in Eq.~\eqref{JointSuccess2}. Using this, the mutual information between Bob and Charlie for the unambiguous communication channel can be written as
\begin{equation}
I_\mathrm{USD}(B:C)=P_\mathrm{ss}H\left( \frac{\eta_{1}(1-q_{1\text{b}})(1-q_{1\text{c}})}{P_\mathrm{ss}} \right).
\label{JointMIBC}
\end{equation}
$I_\mathrm{USD}(B:C)$ is maximized when the information extracted by Bob and Charlie is symmetric, which requires $p_{1b}=p_{1c}$, $p_{2b}=p_{2c}$ and $t=\sqrt{s}$. This can be shown by setting $\frac{\partial}{\partial t}P_{ss}=0$, which leads to $s^{2} q_{1c} = t^{4} q_{1b}$. (Note that the other extremal solution, $P_{ss} = \eta_{2}p_{2b}p_{2c}$, corresponds to the minimum, $I_{USD}\left( B:C \right) = 0$.) Upon inserting the optimum conditions in \eqref{JointMIBC}, we obtain
\begin{equation}\label{eq:MuInfoBCSymmetric}
I_\mathrm{USD}(B:C)=P_\mathrm{ss}H\left( \frac{\eta_{1}(1-q_{1b})^2}{P_\mathrm{ss}} \right),
\end{equation}
where $P_\mathrm{ss}$ is given by Eq.~\eqref{eq:SuccessProbq1b} and the information is symmetrically distributed between Bob and Charlie. Figure~\ref{fig:BCMutualInfoEqual} displays the dependence of $I_\mathrm{USD}(B:C)$ on the parameters of the problem. 

The solid curves in Fig.~\ref{fig:BCMutualInfoEqual} illustrate how $I_\mathrm{USD}$ depends on $q_{1b} (=q_{1c})$ for different values of $s$ for the case of equal priors $\eta_1=\eta_2$. $I_\mathrm{USD}(B:C)$ is optimized by the same solution, $p_{1b}=p_{2b}=1-\sqrt{s}$, as the mutual information between Alice and Bob. It gives a local maximum for the joint probability of success $P_\mathrm{ss}$. Hence, the upper bound of mutual information for equal priors, $\eta_1=\eta_2=1/2$, is
\begin{equation}
I_\mathrm{USD}(B:C)\leq (1-\sqrt{s})^2 \,.
\end{equation}
For the optimal solution, the bit values 0 and 1 occur with the same frequencies in the bit string shared by Bob and Charlie. Obviously, there is no information transmitted through this quantum communication channel at the boundary solutions $q_{1b}=1$ or $q_{1b}=s^2$, where $P_\mathrm{ss}$ can be maximized. It is because at the boundary solution, only one type of bit can be sent and no useful information is effectively communicated through the quantum channel.

\begin{figure}[ht]
\centerline{\setlength{\unitlength}{1pt}
\begin{picture}(230,158)(0,0)
\put(0,0){\includegraphics[width = 220pt]{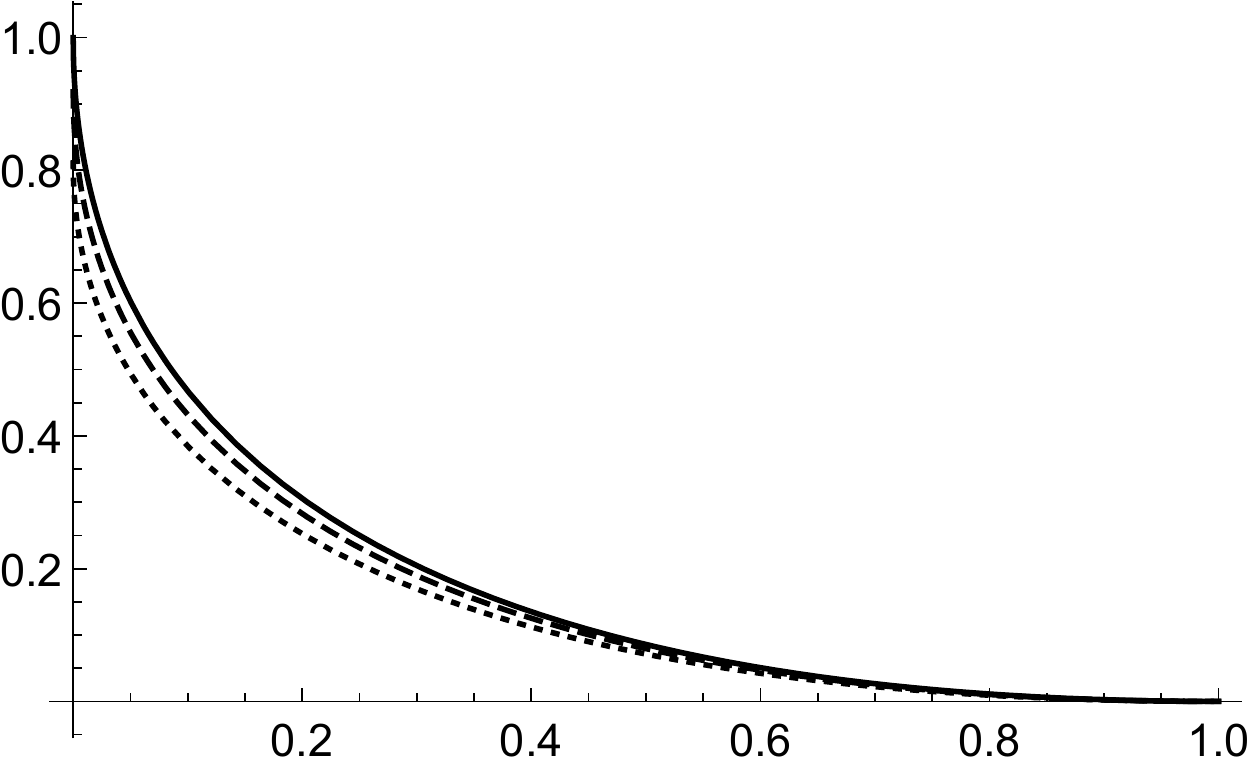}}
\put(85,40){\fbox{\includegraphics[width = 120pt]{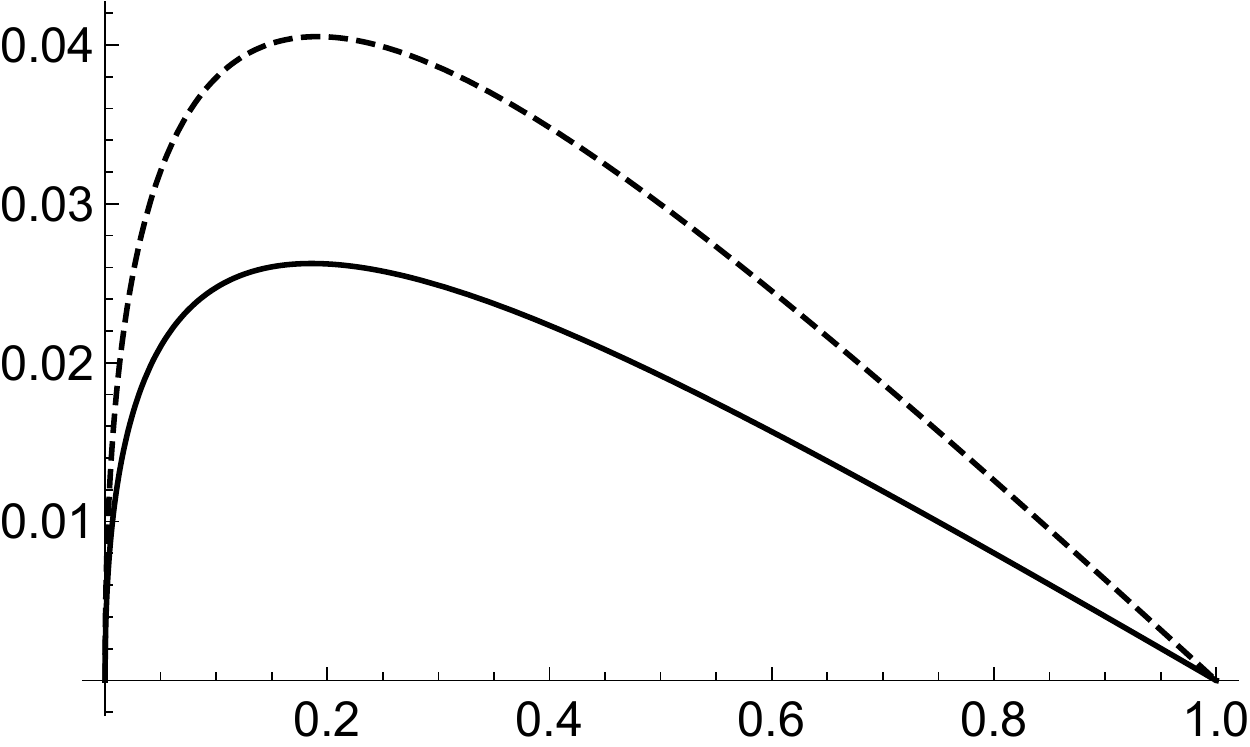}}}
\put(114,-2){\makebox(0,0){$s$}}
\put(150,40){\makebox(0,0)\scriptsize{$s$}}
\put(115,145){\makebox(0,0){Optimal $I_\mathrm{USD}(B:C)$ against $s$}}
\scriptsize{
\put(170,108){\makebox(0,0){$\sqrt{s}-(q_{1b})_\mathrm{opt}$}}
}
\end{picture}}
\caption{Optimal mutual information between Bob and Charlie $I_\mathrm{USD}(B:C)$ as function of the overlap of the signal states $s$. Solid line: $\eta_{1}=1/2$. Dashed line: $\eta_{1}=1/3$. Dotted line: $\eta_{1}=1/4$. The lines are visually indistinguishable from the plots of the approximate solution, $I_\mathrm{USD}(B:C)=(1-\sqrt{s})^2H(\eta_1)$, obtained for $q_{1b}=q_{1c}=\sqrt{s}$. The value of $q_{1b}$, maximizing the mutual information, is $q_{1b}=\sqrt{s}$ for equal priors, and remains close to this value for general priors. The difference $\sqrt{s}-(q_{1b})_\mathrm{opt}$ vs. $s$ is shown in the insert.}
\label{fig:BCMutualInfoMax}
\end{figure}
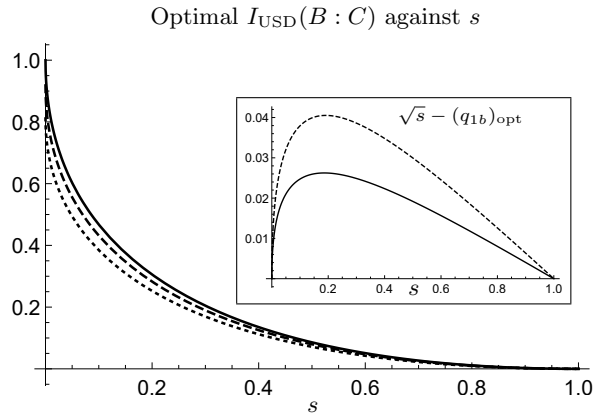

For $\eta_1\neq\eta_2$, the dependence of the mutual information on $q_{1b}$ is very similar the case for equal priors,  Fig. \ref{fig:BCMutualInfoEqual}. However, we no longer have a simple closed analytical upper bound for the mutual information. Instead, we have to rely on numerics. The optimal values of the mutual information $I_\mathrm{USD}(B:C)$ for different prior distributions are shown in Fig.~\ref{fig:BCMutualInfoMax}. 

For $\eta_1= \eta_2$, we had $I_\mathrm{USD}(B:C)\leq(1-\sqrt{s})H(\frac{1}{2})=(1-\sqrt{s})^2$ with the optimum obtained at $q_{1b}=\sqrt{s}$. For $\eta_1\neq\eta_2$, the optimal $I_\mathrm{USD}(B:C)$ is only slightly larger than its value obtained at $q_{1b}=\sqrt{s}$, which is $I_\mathrm{USD}(B:C)=(1-\sqrt{s})^2H(\eta_1)$. Clearly, $I_\mathrm{USD}(B:C)$ and $P_\mathrm{ss}$ are optimized for different values of the parameters.

\subsection{Flip-flop measurement: two- and three-party communication}

Although suboptimal, the flip-flop measurement enables useful and unambiguous information to be transmitted through the communication channel using only von Neumann measurements. The unambiguous mutual information, Eq.~\eqref{eq:MutualInfoUSD2}, depends on the probability of success and the Shannon entropy of the conclusive outcomes. For the flip-flop measurement the mutual information between Alice and Bob can be written as
\begin{eqnarray}
&&I_\mathrm{USD}^{ff}(A:B) = P_s H\left(\frac{\eta_1 c(1-s^2)}{P_s}\right)  \\
&=&[\eta_{1}c+ \eta_{2}( 1 - c)](1 - s^{2})H\!\left( \frac{\eta_{1}c}{ \eta_{1}c + \eta_{2}\left( 1 - c \right)} \right)\!\!. \nonumber
\label{C_Mutual_Info}
\end{eqnarray}
For $c=\eta_{2}$, it reduces to $2\eta_{1}\eta_{2}(1-s^{2})$. For $\eta_1= \eta_2$ it is half the value reached by the POVM measurement. 
For the case of equal priors, $\eta_1=\eta_2=1/2$, the success probability is $P_\mathrm{s}=\frac{1}{2}(1-s^2)$ and the Shannon entropy of the conclusive outcome is $H(c)$. The maximum mutual information is $I_\mathrm{USD}(A:B)=\frac{1}{2}(1-s^2)$ obtained when $c=1/2$. For unequal priors, $\eta_1\neq\eta_2$, the extrema are obtained when the derivative with respect to $c$ is zero, i.e., $\di{c}I_\mathrm{USD}(A:B)=0$. This results in an equation for $c$, 
$\eta_{1}\log\left( \frac{c\eta_{1}}{P_{\text{s}}} \right) = \eta_{2}\log\left( \frac{\eta_{2}\left( 1 - c \right)}{P_{\text{s}}} \right)$,
that can be solved numerically. This equation has only one solution for $0 \geq c \geq1$ which gives the maximum of mutual information.

\begin{figure}[ht]
\centerline{\setlength{\unitlength}{1pt}
\begin{picture}(200,145)(0,0)
\put(0,0){\includegraphics[width = 180pt]{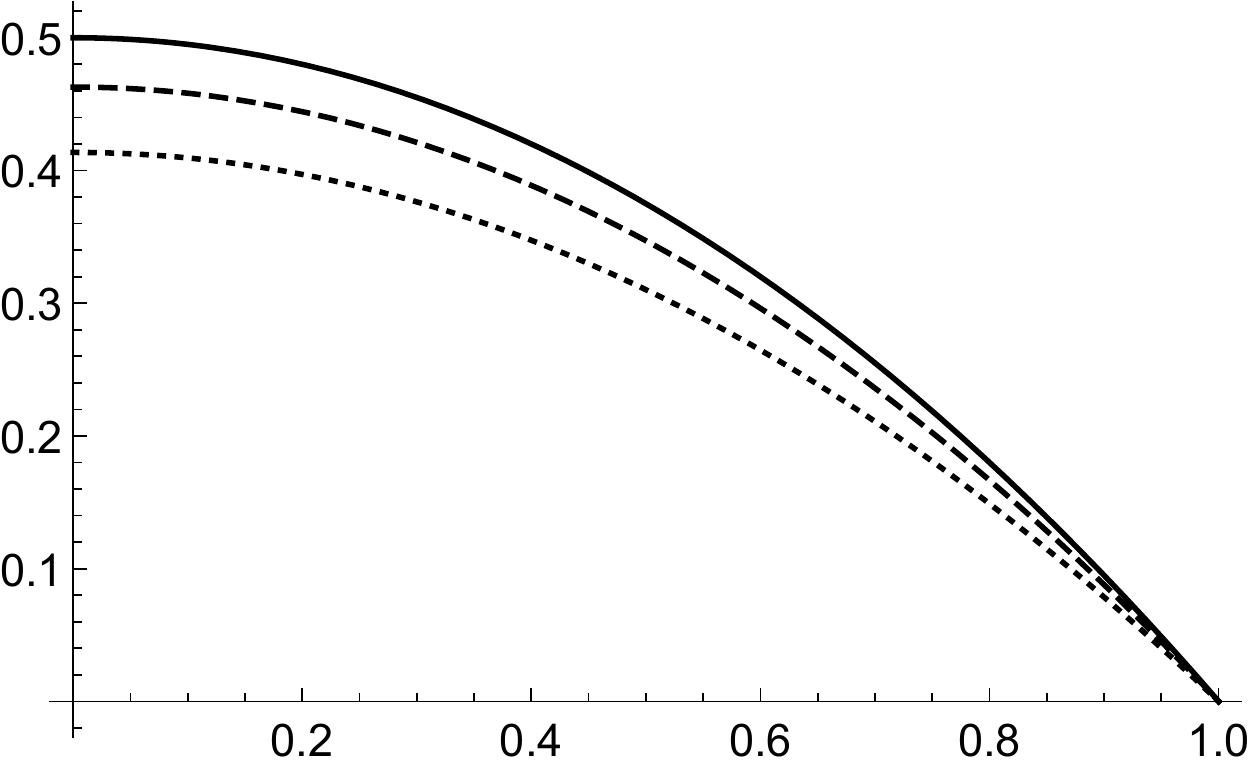}}
\put(95,-2){\makebox(0,0){$s$}}
\put(-28,-15){
\put(115,140){\makebox(0,0){Optimal $I_\mathrm{USD}^{ff}(A:B)$ vs. $s$ for the flip-flop measurement}}
}
\end{picture}}
  \caption{Optimal mutual information, $I_\mathrm{USD}^{ff}(A:B)$, for the unambiguous communication channel between Alice and Bob vs. $s$ using the flip-flop measurement. Solid line: $\eta_{1}=1/2$. Dashed line: $\eta_{1}=1/3$. Dotted line: $\eta_{1}=1/4$.}
\label{fig:MutualInfoFF}
\end{figure}
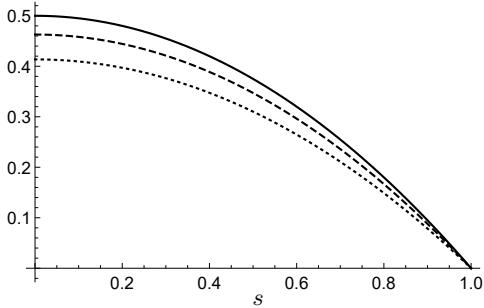

Figure~\ref{fig:MutualInfoFF} shows the maximum $I_\mathrm{USD}^{ff}(A:B)$ of the flip-flop measurement as a function of the overlap of states $s$ for different priors. 

\begin{figure}[ht]
\centerline{\setlength{\unitlength}{1pt}
\begin{picture}(230,158)(0,0)
\put(0,0){\includegraphics[width = 220pt]{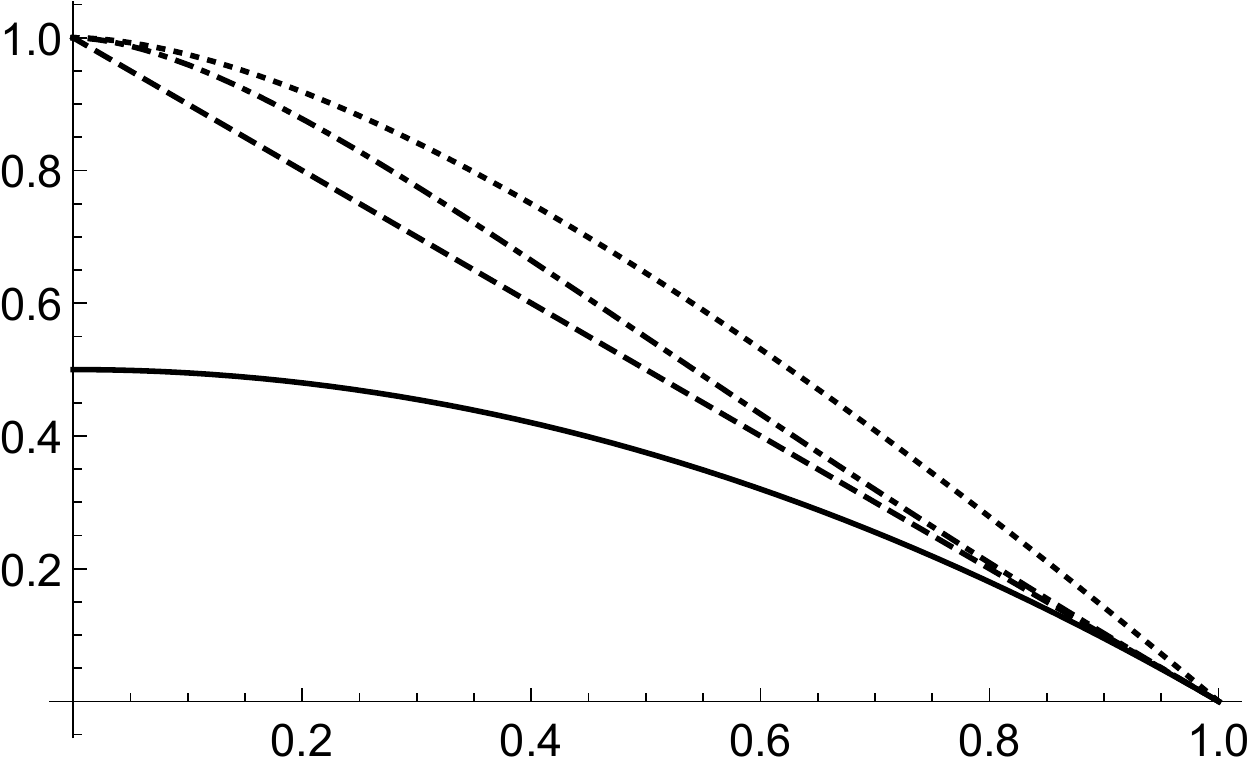}}
\put(114,-2){\makebox(0,0){$s$}}
\put(115,145){\makebox(0,0){Comparing the mutual information for different strategies}}
\end{picture}}  \caption{Mutual information $I(A:B)$ between Alice and Bob as a function of $s$ for different state discrimination strategies and for $\eta_1=\eta_2=1/2$. Dotted line: accessible information $I_\mathrm{acc}(A:B)$ achieved by the Helstrom measurement \cite{Helstrom}, for this communication channel using binary pure state signals. Dot-dashed line: mutual information $I_{\mathrm{boundary}}(A:B)$ for the boundary solutions where one of the two states can be discriminated unambiguously. Dashed line: maximum mutual information for the unambiguous discrimination scheme $I_{\mathrm{USD}}(A:B)$. Solid line: optimal mutual information $I_\mathrm{USD}^{ff}(A:B)$ given by the flip-flop measurement scheme.}
  \label{fig:MutualInfo1Measure}
\end{figure}

Flip-flopping between the two different von Neumann set-ups enables the transmission of unambiguous and useful information between Alice and Bob. The mutual information, however, is bounded by discarding the inconclusive result $\Pi_0$. Even if the signal states are orthogonal, $s=0$, some of the measurement outcomes are wrongly discarded and the maximum mutual information does not reach unity. For equal priors and $s=0$, half of the bits are discarded as inconclusive. For unequal priors, the probability that the outcome is discarded as inconclusive when $s=0$ is $1-P_s=\eta_2+(\eta_1-\eta_2)(1-c)$. The mutual information $I_\mathrm{USD}(A:B)$ for the flip-flop measurement, the accessible information and the optimal unambiguous discrimination strategy are compared in Fig.~\ref{fig:MutualInfo1Measure}.

Clearly, the Helstrom measurement yields the highest information gain, as it should \cite{Keil2008}.

In order to complete the study of the three-party communication scheme using the flip-flop measurement, we now address the mutual information between Bob and Charlie. Using Eqs. \eqref{Sequential FFM q1}-\eqref{FF1}, $I_\mathrm{USD}^{ff}(B:C)$ becomes
\begin{eqnarray}
	I_\mathrm{USD}^{ff}(B:C) & = & (1 - \frac{s^{2}}{t^{2}})\left( 1 - t^{2} \right)\left( \eta_{1}c^{2} + \eta_{2}\left( 1-c \right)^{2} \right) \nonumber \\
	&& *H\left( \frac{\eta_{1}c^{2}}{\eta_{1}c^{2} + \eta_{2}\left( 1-c \right)^{2}} \right).
	\label{SFFMI}
\end{eqnarray}
This is clearly minimum (=0) when $c=0$ or $c=1$, i.e., at the boundaries of the allowed range for the flipping rate. Thus, no unambiguous information is transferred between Bob and Charlie in this case. 

With the optimal $c=\eta_{2}$ and $t^{2}=s$, however, we obtain
\begin{eqnarray}
	I_\mathrm{USD}^{ff}(B:C)  =  \eta_{1} \eta_{2}(1 - s)^{2} H\left( \eta_{2} \right) \,,
	\label{SFFMI2}
\end{eqnarray}
for the mutual information that is useful for establishing a quantum communication channel between Bob and Charlie. It is always less than the POVM optimum.

\section{Discussion and conclusion}
In this paper, we made a number of important additions to the theory of both standard and sequential unambiguous discrimination. In order to make the paper self-consistent, we started with a brief overview of the sequential unambiguous scheme \cite{Bergou2013}, where the optimal joint probability of success of the subsequent observers was obtained for the case when the possible initial states of the system were prepared equally likely. After these preliminaries, we first suggested a multiparty communication protocol achieved via a single qubit,  based on our scheme. Next, we introduced a novel scheme in which, in addition to optimizing the joint success probability, the joint probability of failure is also minimized and gave a fully analytical solution for this strategy. Then, for the scheme when only the joint probability of success is optimized without minimizing the joint probability of failure, we presented analytical and numerical optimal solutions, with particular attention to local and global optima. Perhaps most importantly, we worked out the theory of mutual information for the unambiguous discrimination strategy, in order to fully take into consideration the restrictions on the information gain inherent in this strategy, and showed that the mutual information conditioned on success is the quantity consistent with the unambiguous discrimination scheme. We applied these considerations to the calculation of mutual information for both the standard and the sequential schemes and showed that the boundary solutions \cite{Pang2013}, carry no useful information for quantum communication. Therefore, we introduced  the so-called flip-flop measurement \cite{Bennett92}, to salvage the boundary solution and make it useful for quantum communication. Finally, we showed that the maximum of the mutual information conditioned on success coincides with the maximum of the success probability of unambiguous discrimination, making it the proper measure of information gain for the scheme that employs the sequential unambiguous discrimination strategy as a multiparty communication channel.

In addition, our work has two quite general messages.  The first is concerned with the statement that can be found in many textbooks on quantum mechanics, the so-called collapse postulate: the state of the system right after a measurement is performed on it is one of the eigenstates of the observable that has been measured.  Therefore, a subsequent measurement will detect this state and yield no information about the state of the system before the measurement.  Our work shows that, at least probabilistically, one can get even unambiguous information about the initial preparation of the system.  The second, closely related, issue is concerned with the no-broadcasting theorem: a single qubit cannot be broadcast to more than one observer.  Again, what we show here is that, at least probabilistically, one can get around this theorem and more than one observer can get information about the initial preparation of the qubit.  We fully expect that these general messages will trigger further investigations along the lines studied here.


\end{document}